\documentstyle[11pt,epsfig,epsf]{article}
\newcommand\C{\v Cerenkov }
\def\jnfont{\rm}
\def\NPB#1,{{\jnfont Nucl.\ Phys.\ }{\bf B#1},}
\def\PLB#1,{{\jnfont Phys.\ Lett.\ B }{\bf #1},}
\def\PRD#1,{{\jnfont Phys.\ Rev.\ D }{\bf #1},}
\def\PRL#1,{{\jnfont Phys.\ Rev.\ Lett.\ }{\bf #1},}
\def\ZPC#1,{{\jnfont Z.~Phys.\ C }{\bf #1},}
\def\ETslash{\not{\hbox{\kern-4pt $E_T$}}}
\let\to=\rightarrow
\begin{document}
\thispagestyle{empty}
\begin{flushright}
VLBL Study Group--H2B-1\\
IHEP-EP-2001-01\\
AS-ITP-01-004\\
Ames-HEP 01-01\\
\today 
\end{flushright}
\vskip 3ex
\begin{center}
Report of a Study on
\vskip 2ex
\vskip 2ex
{\Large H2B\\
\vskip 2ex
Prospect of a very long baseline neutrino oscillation experiment \\
\vskip 1ex
HIPA to Beijing}
\end{center}
\vskip 4ex
\begin{center}
Hesheng Chen, Linkai Ding, Jingtang He, Haohuai Kuang, Yusheng Lu, 
Yuqian Ma, Lianyou Shan\\ 
Changquan Shen, Yifang Wang, Changgen Yang, Xinmin Zhang and Qingqi Zhu\\
 
\vskip 1ex
{\bf Institute of High Energy Physics, CAS, Beijing}\\
\vskip 2ex
Chengrui Qing, Zhaohua Xiong, Jin Min Yang and Zhaoxi Zhang \\
\vskip 1ex
{\bf Institute of Theoretical Physics, CAS, Beijing}\\
\vskip 2ex
Jiaer Chen and Yanlin Ye \\
\vskip 1ex
{\bf Institute of Heavy Ion Physics, Peking University, Beijing}
\vskip 2ex
S.C. Lee and H.T. Wong \\
\vskip 1ex
{\bf Institute of Physics, AS, Taipei} \\
\vskip 2ex
Kerry Whisnant and Bing-Lin Young\\
\vskip 1ex
{\bf Department of Physics and Astronomy, Iowa State University, Ames}\\




\end{center}

\newpage

\tableofcontents

\addtocontents{loi6.toc}{References}

\newpage

\noindent

\section{Introduction}

\vskip 2ex

Neutrino oscillations, if neutrinos are massive, were first proposed 
by Pontecorvo \cite{Pontecorvo} and by Maki, et al., \cite{Maki},
respectively, in the late 1950's and early 1960's. It is to date
the only experimental indication of physics beyond the standard model
(SM) that may provide a starting point to further expand the horizon
of our most basic knowledge of nature. Historically neutrinos have
played mostly passive roles in particle physics research.  However,
the observation of neutrino oscillations by the Super-Kamiokande
experiment~\cite{superK}, which are corroborated by various other
experiments, changed significantly the role of neutrinos and it has
very profound implications for particle physics, astrophysics and
cosmology.  Neutrinos are the end product of the decays of almost all 
elementary particles, observed or proposed. They fill the universe as
a relic background radiation similar to the photon as a finger print
of the history of the universe.  Now they are taking the center stage
in particle physics research and their heightened fascination by
physicists has just began.

Since it was proposed more than thirty years ago, the standard model of 
electroweak and strong interactions (SM) has been firmly established as
the fundamental theoretical framework of elementary particles.  Its   
applications straddles eleven orders of magnitude in energy range, from
the atomic parity violation of 1 eV to the W, Z and top quark physics 
of hundreds of GeV.  All high energy phenomenologies that have been
scrutinized by experiments are found to agree with the predictions of
the SM.  The only missing piece is the  Higgs particle.  Even for that, 
a tentative gleam of its existence from the last LEP data \cite{LEPHiggs}
indicated that it may be what is expected from the SM.  In reaching the
present status of the SM, the IHEP's BEPC program has made the important
contribution in establishing the lepton universality which is a corner
stone of the SM. In this ``business as usual'' state of affairs, evidence
for neutrino oscillation and hence massive neutrinos gives light to the
possibility of a new point of departure for post SM physics.

From the available neutrino oscillation data, we can already see some rich 
features of the neutrino sector.  In analogy to the mass spectra of the
quarks and charged leptons, there is a hierarchical structure in the mass
square differences of the neutrinos.  But in a departure from the feature
of the quark sector, the mixing angles of the neutrinos are large, some
probably maximal.  Another intriguing aspect of the neutrino oscillations
is that they are manifested macroscopically at terrestrial and solar
distances.  The physics of neutrino oscillation can be treated effectively
by simple quantum mechanics.  When finally confirmed, it will be an unique
and elegant illustration of the effect of quantum interference at macroscopic
scales.

Although the existing data offer strong indication of neutrino oscillations,
the oscillation parameters have not been determined with sufficient 
accuracy.  The unique signature of the flavor transmutation, i.e., the 
appearance of a flavor different from the original one, has not been
convincingly observed.  In the ongoing and next generation neutrino 
oscillation experiments under construction, some of the parameters will be 
probed with greater accuracy.  But some other parameters may not yet be
accessible.  Hence more experiments designed to look for the missing
information and to probe the known parameters with even greater accuracy
are necessary.  

One possible implication of the macroscopic manifestation of the neutrino 
oscillation is that the effect of the lepton flavor mixing may be very
small in microscopic distances.  Hence there are possibly very little
observable effects to the conventional high energy physics parameters
measured at microscopic distances, such as the charge lepton flavor
changing neutral current \cite{BDWY}.  This provides additional motivation
for pursuing refined neutrino oscillation experiments so that the full
implication of neutrino flavor mixing can be thoroughly investigated and
understood, although the search for the effects at microscopic distances
should be pursued. 

This report summarizes the preliminary results of our study on a
very long baseline (LBL) neutrino oscillation experiment, which uses the
intensive conventional neutrino beam from the High Intensity Proton
Accelerator (HIPA)~\cite{hipa}, Japan, or the beam of a possible neutrino factory,
delivered to a detector located in
Beijing, China, tentatively called the Beijing Astrophysics and Neutrino
Detector (BAND). HIPA to BAND (H2B) has a distance of 2100 km, which is eight
times longer than the currently online LBL experiment, K2K, and three
times longer than all of the three LBL experiments in construction:
MINOS, ICARUS and OPERA.  This very long baseline program has the
distance to neutrino energy optimal for the atmospheric neutrino
oscillation scale.  It enables us to investigate some of the parameters
that are not easily accessible by the above mentioned on-going and approved
experiments.  For a neutrino factory located in North America or Europe
the distance to BAND is even much longer.

By the time this experiment becomes online some of the basic parameters
related to the solar and atmospheric neutrino oscillations will be
accurately determined.  BAND will be in position to carry out a refined
investigation of the neutrino oscillation 
 parameters, such as the mixing parameters
$\sin^22\theta_{13}$, the matter effect and the sign of the dominant
neutrino mass-squared difference.  Additional important avenues of
investigation include the effect of CP violation, some of the astrophysics
measurements, the neutral current reactions that have scanty experimental
information available to date, etc.  The recent evidence of direct CP
violation in the hadron sector \cite{hadronCP} (and therefore evidence
of non-vanishing phase in the CKM matrix) suggests that a non-vanishing
phase angle may also exist in the lepton sector.  Furthermore, the large
mixing angles of the neutrinos makes it plausible to expect that the CP
phase angle is likely large in the lepton sector.  

This report is motivated by the discussions took placed at several
workshops held at IHEP during the past two years.  Several Physicists
from Japan and USA joined their Chinese colleagues in the workshop
held on May 25, 2000~\cite{workshop}.  As an outcome, an agreement was made to form
two study groups, both consisting of experimentalists and theorists,
to make independent studies of the physics potential of H2B.
One group,
coordinated by Dr. Kaoru Hagiwara, is made of physicists from KEK
and the University of Kyoto and the other group, which is responsible
for the present study report, includes physicists from China and USA.
We believe our study so far and that
of the KEK-Kyoto group~\cite{kekkyoto} show
that the physics of H2B is compelling and merits serious considerations
for this very long baseline neutrino oscillation experiment.  Further
studies using more realistic neutrino beam parameters with detailed 
detector design are called for.  

We envisage that the experimental programs outlined in this report will
take a significant lead time to be fully implemented.  The long lead
time required is partly due to the R\&D efforts of BAND but mainly
constrained by the timeline of the neutrino beam, such as that from
HIPA or a neutrino factory.  We should also note that the long lead
time is characteristic of modern large high energy physics experimental
programs.  Anticipating the long lead time, BAND will be constructed in
the flexible modular approach and designed to do astrophysics experiment
in the first stage of the program. In the intervening time, BAND will be
thoroughly tested, enlarged and improved to be ready for the exciting
physics of a LBL neutrino experiment.

In Sec. 2 we summarize the current experimental status of neutrino 
oscillations and the expectations of the new generation of experiments
online and in construction. In Sec. 3 we summarize the theory of neutrino 
oscillations and list some of useful formulae.  Section 4 outlines the
fundamentals of the long baseline neutrino oscillation experiment and
presents the result of a preliminary study of the physics expectations
of H2B.  Sec. 5 presents a preliminary design of a possible BAND detector
as a concrete description of the physics results reachable at H2B.

A number of topics which are not covered by the present report will be
the focus of the future studies, which involve a refined study of
possible detector(s), study of other physics opportunities including
those listed in Sec. 5.3, oscillation physics at even long oscillation
distances that may involve BAND, and the comparison of H2B with other
LBL experiments in construction.  
Topics such as the near detector, the neutrino beam    
design, and the choice of a far detector are expected
to be the joint effort of a full collaboration with the
participation of many physicists world wide, including
the present study group, the critical participation 
from Japan, and from institutions of many other regions
and countries.

\newpage
\noindent

\section{Current experimental status of neutrino oscillations}

In the following we review briefly the present experiments and the relevant 
data.  For more details we refer to review articles available in the 
literature; we list a few in Ref. \cite{expreview}.  We do not claim to 
be exhaustive and apologize for the omission of any articles and experimental 
data.  For an exhaustive list of neutrino experiments 
see the neutrino web-site in Ref. \cite{website}.
Since the interpretation of the data are mostly in the two-flavor scenario,
we will first summarize the formalism of two-flavor oscillations in vacuum.

\subsection{Two-flavor neutrino oscillations}

We begin in the lepton flavor framework in which the charged lepton mass 
matrix is diagonalized.
Define the two-flavor eigenstates $\nu_\alpha$ and $\nu_\beta$, and their mass 
eigenstates, $\nu_1$ and $\nu_2$ of masses $m_1$ and $m_2$, respectively. 
The flavor states and mass eigenstates are related by a mixing matrix $U$, 
which is an orthogonal transformation in two dimensions:
\begin{equation}
\left( \begin{array}{l} \nu_\alpha  \\  \nu_\beta \end{array} \right)
  = U \left( \begin{array}{l} \nu_1 \\ \nu_2 \end{array} \right),~~~~~
  U= \left( \begin{array}{rr} \cos\theta  &  \sin\theta  \\
                            -\sin\theta & \cos\theta \end{array} \right)
\end{equation}
where $\theta$ is the {\bf mixing angle}.  
The states are orthonormalized within their own spaces, i,e.,
\begin{equation}
<\nu_j|\nu_k> = \delta_{jk},~~~j,k=\alpha, \beta;~ {\rm or}~ 1, 2.
\end{equation}
It should be noted that in an experiment, neutrinos are always produced as
flavor eigenstates.\\

The time evolution of a flavor
state can be simply expressed in terms of the time evolution of the mass
eigenstates which enter into the flavor state at $t=0$, 
\begin{equation}
\left( \begin{array}{l} \nu_\alpha(t) \\ \nu_\beta(t) \end{array} \right)
  = U \left( \begin{array}{l} \nu_1(t) \\ \nu_2(t) \end{array} \right)
  = U \left( \begin{array}{l} e^{-iE_1 t}\nu_1 \\ e^{-iE_2 t}\nu_2
             \end{array} \right)
  = U \left( \begin{array}{ll} e^{-iE_1 t} &  0  \\
                        0  & e^{-iE_2 t} \end{array} \right)U^\dagger 
      \left( \begin{array}{l} \nu_\alpha \\ \nu_\beta \end{array} \right),
\end{equation}

Suppose the neutrino flavor state $\nu_\alpha$ is produced, then at time t
we have
\begin{equation}
\nu_\alpha(t) =
  (\cos^2\theta e^{-iE_1t} + \sin^2\theta e^{-iE_2t})\nu_\alpha
  + \cos\theta \sin\theta (-e^{-iE_1t} + e^{-iE_2t})\nu_\beta.
\end{equation}
The probability of finding the original flavor, 
referred to as the {\bf survival probability}, is
\begin{eqnarray}
P_{\nu_\alpha\to \nu_\alpha}=|<\nu_\alpha|\nu_\alpha(t)>|^2
       &  = & 1-\sin^2(2\theta)\sin^2\left({E_2 -E_1\over 2}t\right)
              \nonumber \\
       &  = & 1 -\sin^2(2\theta)
    \sin^2\left(1.267\Delta m^2({\rm eV^2}){L({\rm km})\over
                                  E_\nu({\rm GeV})}\right)
\end{eqnarray}
and the probability of finding the other flavor, called the
{\bf appearance probability} is
\begin{equation}
P_{\nu_\alpha\to \nu_\beta}= \sin^2(2\theta) 
    \sin^2\left(1.267\Delta m^2({\rm eV^2}){L({\rm km})\over
                                  E_\nu({\rm GeV})}\right).
\end{equation}
Here we take the approximation 
$E_j \approx |\vec{p}| + m_j^2/2E_\nu$ and denote
$\Delta m^2 = m^2_2 -m^2_1$.

The characteristic behavior of this expression as a function of 
$\sin^2(2\theta)$ and $\Delta{m}^2$ is the following: for large
$\Delta m^2$,
the argument of the sine function is large and
hence oscillates rapidly in even a very small energy range.
The energy average of the sine
function involved becomes ${1\over 2}$, hence we have
\begin{equation}
\sin^2(2\theta) \approx 2 P_{\nu_\alpha \to \nu_\beta}.
\end{equation}
For small $P_{\nu_\alpha\to \nu_\beta}$ and near $\sin^2(2\theta)=1$ the 
curve behaves like

\begin{equation}
\Delta{m}^2\approx
             {\sqrt{P_{\nu_\alpha\to \nu_\beta}} \over 1.267 L/E_\nu}
\end{equation}
Hence the effective probe in $\Delta{m}^2$
lies in the  region
\begin{equation}
{\sqrt{P_{\rm min}}\over 1.267 L/E_\nu} < \Delta{m}^2
    < {\pi\over 1.267 L/E_\nu}
\end{equation}

This feature is true even for cases of more than two flavors of neutrinos.
If the different mass squared differences are well separated, the effective
probe of each mass squared is in the region of $L/E_\nu$ which satisfies
the above relation. 

\vskip 2ex

\noindent

\subsection{Evidence of neutrino oscillations}

\begin{figure}[htbp!]
\begin{center}
\mbox{\epsfig{file=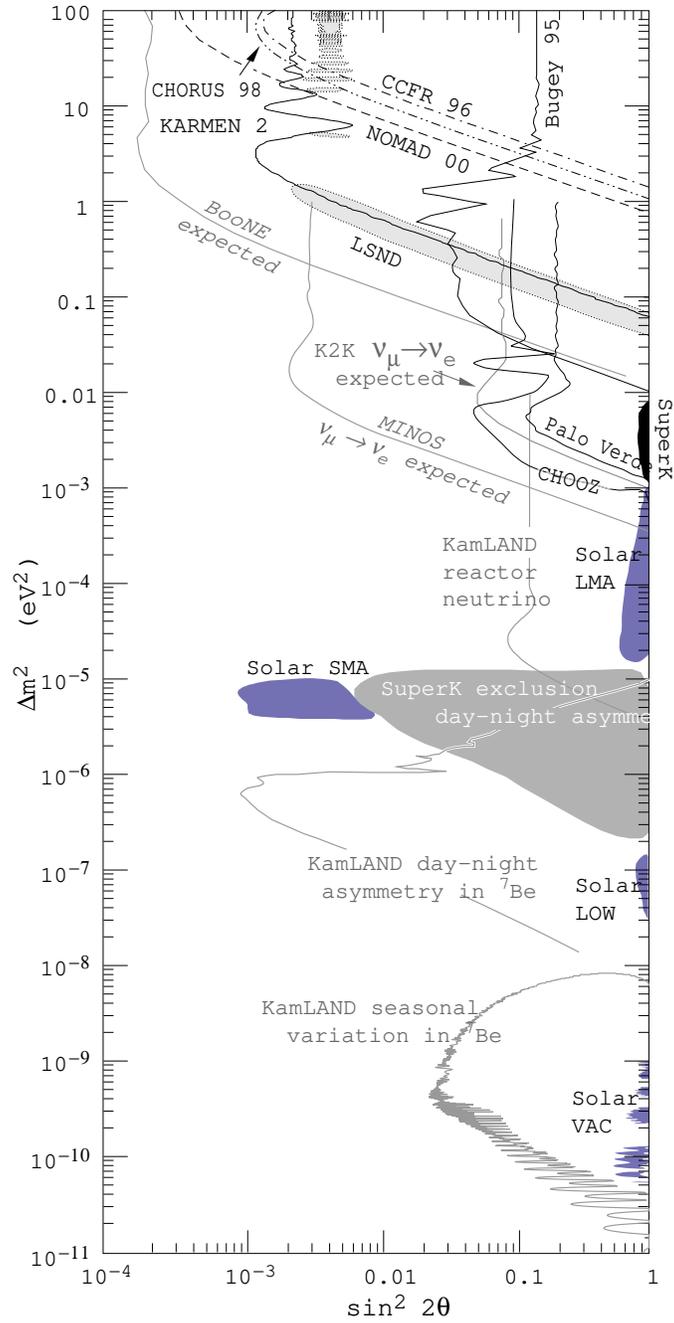,width=10cm,clip=}}
\caption{Summary of current experimental results on neutrino oscillations.
Filled 
areas are allowed regions from the evidence of neutrino oscillations.
Excluded regions are up and right of curves. Sensitivity
curves of future experiments are also shown.
}
\label{fig:osc_results}
\end{center}
\end{figure}

The most convincing evidence of neutrino oscillations is given by the
Super-Kamiokande (Super-K) experiment on 
atmospheric neutrinos~\cite{superK}. The solar neutrino data provide 
the corroborative evidence of the oscillatory scenario.  The reactor 
experiments have not seen any evidence of oscillations, but they provide 
strong constraints on some of the allowed regions of the oscillation 
parameters and are complementary to the atmospheric and solar data.  The 
accelerator based short baseline experiments, LSND and KARMEN, 
are so far inconclusive.  Fig.~\ref{fig:osc_results} summarizes the 
pertinent results of all experiments to date. The filled 
areas are allowed $\sin^2(2\theta)-\Delta{m}^2$ regions that provide the 
evidences of neutrino oscillations. Excluded regions lie above and to the 
right of the corresponding curves shown. Sensitivity
curves of some of the future experiments are also shown.

\subsubsection{Atmospheric neutrinos}

The atmospheric neutrino data are the so-called smoking gun of 
neutrino oscillations.  They offer the strongest evidence of neutrino
oscillations to date.  
The first indication of the ``atmospheric neutrino anomaly'' was observed 
in the 1960's by experiments in South Africa and India and was confirmed 
during the 1980's to 1990's by IMB, Kamiokande, FREJUS, NUSEX, Soudan and
MACRO experiments.  The data are usually analyzed in terms of the double 
ratio, namely the observed over 
expected ratios of muon to electron neutrinos, to reduce the systematic
error.  Such ratio of ratios is measured to be about $0.60\pm 0.06$ instead of 
1 as expected, indicating neutrino oscillations.
The breakthrough came in the 1990's when the Super-Kamiokande experiment
went online. With a much larger statistical sample and broad energy range,
they observed the following effects specific to neutrino oscillations~\cite{superK}:

\vskip 1ex

{\bf Zenith angle dependence}: The down-ward($\theta=0^\circ$) and 
upward($\theta=180^\circ$) going neutrinos travel 
a distance varying from L=20 km to L=13000 km. The observation of 
zenith angle dependence of muon neutrino events consistent with the
expected variation of oscillation probability as a function of L.
For electron neutrinos, such variation was not observed and it is consistent
with reactor neutrino experiments which found null oscillations of $\bar\nu_e$
~\cite{chooz,pv}.  Such zenith angle dependence of $\nu_\mu$ 
has also been observed later by the Soudan 2 and the MACRO experiments.

\vskip 1ex

{\bf Energy and L/E dependence}: Muon neutrino events as a function of the
neutrino energy
and L/E have been observed and they are consistent with expected variation of
neutrino oscillation probability. Again, electron neutrinos have 
no such variation.

\vskip 1ex

{\bf East-west anisotropy}: 
The Earth's magnetic field, which deflects charged particles traveling in 
the atmosphere, affects the cosmic charge particles moving eastward 
differently from those moving westward. The neutrino flux model 
which takes into account the effect of the Earth's magnetic field, 
is consistent with observed east-west anisotropy for horizontal 
going $\mu$-like and e-like event, supporting
the neutrino oscillation interpretation\cite{east-west}.

\vskip 2ex

The Super-K data allows for the $\nu_\mu \to \nu_\tau$ oscillation 
in a region shown in Fig.~\ref{fig:osc_results} at 90\% CL. It is also
reported recently~\cite{sterile} that the muon neutrino to a sterile 
neutrino, $\nu_\mu \to \nu_s$, as a dominant
oscillation mode is excluded at the 99\% confidence level (CL). Another exotic
mechanism, i.e., the neutrino decay, although not completely excluded, is not
a satisfactory explanation for the atmospheric neutrino data ~\cite{decay}.

\vskip 2ex

\subsubsection{Solar neutrinos}

The ``solar neutrino deficit'' was first detected in 1968~\cite{Davis} by 
the Homestake experiment and later confirmed by the Kamiokande, GALLEX,
SAGE and Super-K experiments. The observed neutrino flux is only
1/3-1/2 of what was expected by the Standard Solar Model (SSM)
depending on the neutrino energy. 
During the last decade,
the understanding of the solar structure has been significantly improved, 
facilitating the analysis and interpretation of the solar neutrino 
data~\cite{solarwebsite}.

The only plausible explanation for this deficit is 
electron neutrino oscillation.
Fig.~\ref{fig:osc_results} shows 4 allowed regions for 
$\nu_e$ oscillation into $\nu_\mu$ or $\nu_\tau$
in the 
$\Delta{m}^2-\sin^2(2\theta)$ space from results of all solar neutrino 
experiments:
\vskip 1ex 

{\bf Vacuum long wavelength oscillation solution} (VAC): This solution assumes
no matter effect and is also known as the "{\bf just so}" solution.  The best
fit gives a very small squared mass difference and nearly maximal mixing:
\begin{equation}
\Delta{m}^2 ~~\sim~~ 8\times 10^{-11}~{\rm eV}^2,
                       ~~~~\sin^2(2\theta)~~\sim~~ 0.8.
\end{equation}
In this solution the Earth-Sun distance is approximately half of an
oscillation 
length, $L_{osc} = 4\pi E_\nu/\Delta{m}^2$, which implies a significant 
dependence of the $\nu_e$ survival probability on the energy of the neutrino.

\vskip 1ex

{\bf Small mixing angle MSW solution} (SMA):  In this solution the
matter effect is included.  The best fit of the data prior
to the most recent Super-K data gives:
\begin{equation}
\Delta{m}^2 \approx 5\times 10^{-5}~{\rm eV}^2,~~~~
                \sin^2(2\theta)\approx 10^{-3}-10^{-2}.
\end{equation}
\vskip 1ex

{\bf Large mixing angle MSW solution} (LMA): Again the matter effect is included
and  the
allowed parameters are:
\begin{equation}
\Delta{m}^2  \approx 2\times 10^{-5}~{\rm eV^2},
      ~~~~\sin^2(2\theta) \approx 0.8.
\end{equation}
\vskip 1ex

{\bf Low mass solution} (LOW): This is similar to the LMA case but
with a much lower squared mass difference.  The parameters are: 
\begin{equation}
\Delta{m}^2 \approx 8\times 10^{-8}~{\rm eV^2},~~~~\sin^2(2\theta) \approx 1.
\end{equation}


For sterile neutrinos, only VAC and SMA solutions are allowed at 99\% CL
with similar parameters given above. However, 
recent Super-K results~\cite{sknew}
have ruled out both the sterile neutrinos and the VAC and SMA solutions at 95\% CL. 
The LMA and LOW are 
viable with the LMA being the favorable solution at 90\% CL.

It should be remarked that the data do not rule out the oscillation of
$\nu_e$ into two or more flavors with comparable probabilities. Furthermore,
some analysis concluded that the solar 
neutrino data is still ambiguous and the SMA may still be viable\cite{Smirnov}.  
Future data from SNO and
BOREXINO as well as reactor data from KamLAND will be decisive for the
determination of the solar neutrino oscillation parameters. 
Various possibilities of checking the solar neutrino problem can be
found in Ref. \cite{Langacker}.

\vskip 2ex

\noindent

\subsubsection{Neutrino oscillation searches at nuclear reactors}

Nuclear reactors are an intense and stable low energy source of
$\bar{\nu}_e$, which is complementary to the solar neutrinos($\nu_e$).  
The beam emitted from the reactor is isotropic with an average 
energy around 3 MeV and an maximum energy of 10 MeV.  The $\bar{\nu}_e$
flux is known within 2.7\% when the reactor parameters are 
given.  (For references, see Ref. \cite{wangkam}.) 

The two reactor neutrino experiments CHOOZ~\cite{chooz} and Palo
Verde~\cite{pv} both have a baseline of about 1 km and were recently
completed.  Null oscillation results have been reported and they excluded 
$\bar\nu_e \rightarrow \bar\nu_x$ oscillations in a region as shown in
Fig.~\ref{fig:osc_results} at 90\% CL.
An important conclusion from the data of these experimental is that
the atmospheric neutrino oscillation is not predominantly caused by the
$\bar{\nu}_\mu \to \bar{\nu}_e$. 

The future long baseline reactor neutrino experiment, KamLAND, may find the
smoking gun for the solar neutrino problem.  The detector consists of 1 kt
of liquid scintillator and can detect $\bar{\nu}_e$ from all nuclear
reactors in Japan with an event rate of 2/day.  The characteristic baseline
is 160 km and its sensitivity curve is shown in Fig.~\ref{fig:osc_results}.
If the solution to the solar neutrino deficit turns out to be LMA,
KamLAND will be able to see the effect of neutrino oscillations, similar
to the solar neutrino deficit. 

\vskip 2ex

\noindent

\subsubsection{Short baseline accelerator neutrino oscillation experiments}

The accelerator neutrino experiments discussed below are low energy
short baseline experiments which search for the oscillation
$\nu_\mu(\bar\nu_\mu)\to \nu_e(\bar\nu_e)$.  The results came from two
experiments, LSND  and 
KARMEN.
The LSND experiment used a scintillator and Cherenkov detector
with $L= 29$ m and average neutrino energy $E_\nu = 110$ MeV.  It 
has the following results~\cite{LSND} before its completion in 1998: 

\vskip 1ex

{\bf Search for the $\bar{\nu}_\mu \to \bar{\nu}_e$ oscillation}:

The $\bar\nu_\mu$ beam of LSND is produced from $\mu^+$ decay at rest.
A total of 22 $\bar\nu_e$ events are observed via the reaction 
$\bar\nu_e P\to e^+n$, while only $4.6\pm 0.6$ events are expected to 
come from the background.
A fit to the $e^+$ energy spectrum yields an excess of 
$51.8^{+18.7}_{-16.9}\pm 8.0$ events, corresponding to an oscillation
probability of $(3.1^{+1.1}_{-1.0}\pm 0.5)\times 10^{-3}$ if interpreted as 
neutrino oscillations.  The allowed region is shown in
Fig.~\ref{fig:osc_results}.

\vskip 2ex

{\bf Search for the $\nu_\mu\to \nu_e$ oscillation}:

In their later data sample, LSND has observed 40 events in 
$\nu_\mu\to \nu_e$ by using $\nu_\mu$ from $\pi^+$ decay in flight.
The $\nu_e$ signal is detected via the charged current reaction 
$\nu_eC\to e^-X$.
The expected  backgrounds consist of
$12.3\pm 0.9$ events from cosmic ray and $9.6\pm 1.9$ from other neutrino
induced processes.  Interpreted as neutrino
oscillation, the excess of $18.1\pm 6.6$ events correspond to
a $\nu_\mu\to \nu_e$ probability of $(2.6\pm 1.0)\times 10^{-3}$ which
is consistent with the $\bar{\nu}_\mu\to \bar{\nu}_e$ oscillation
probability given above.  Since this is a $2.6\sigma$ effect, confirmation 
with higher statistics is necessary.
\vskip 2ex

The KARMEN experiment uses a liquid scintillator detector with 
$L=17$~m at a comparable neutrino energy.  
KARMEN has 
not observed any oscillation signals. Its excluded region is also shown in 
Fig.~\ref{fig:osc_results}.  
The data taking by KARMEN will be finished in 2001. Its sensitivity to 
$\bar{\nu}_\mu\to \bar{\nu}_e$ will be increased by a factor 1.7,  but 
still cannot cover the whole LSND allowed region.  

The Fermilab Mini-BooNE  experiment
which will be online in 2001, can cover the whole LSND allowed region. 
The Mini-BooNE $\nu_\mu$ beam, which has a broad band of 0.3-2 GeV and 
$\nu_e$ contamination of about 0.3\%,
 is produced by the Fermilab 8 GeV, high 
intensity Booster Synchrotron proton beam.  The 445 ton mineral oil
detector is located 500 m from the 
neutrino source.  The expected $\nu_\mu\to \nu_e$ oscillation suggested by 
the LSND data will 
produce 1500 excess events of the electron type which is many $\sigma$'s 
away from the statistical error of the experiment.  The sensitivity curve 
is reported in Fig.~\ref{fig:osc_results}.

%
%
%
We summarize in 
Table~\ref{tab:expsummary} the current status on the determination of the
$\Delta{m}^2$ and $\sin^2(2\theta)$ parameters in the various types of
experiments.  Some of the most recent Super-K data are included.  It should
be emphasized again that the best values of the oscillation parameters
are obtained assuming two-neutrino mixing.

\begin{table}[htbp]
\begin{center}
\begin{tabular}{llll}
Exp       & Sources             & Phenomenon   & Recent results/remark  \\
\hline \hline
Reactor   & $\bar\nu_e$       & $\bar\nu_e$ survival  
          & No oscill observed, rule out regions \\  
    &  &  & Expected future data from:\\
    &  &  & KamLAND \\ \hline
Solar     & $\nu_e$             & $\nu_e$ deficit            
          & Regions allowed: LMA \& LOW \& SMA \& VAC\\
    &  &  & Expected future data from:  \\
    &  &  & Super-K, SNO, BOREXINO      \\ \hline
Atmospheric  & $\nu_\mu,\nu_e $ & $\nu_\mu$ deficit
          & Dominant (@99\% CL): \\
smoking gun  &  &  &  $\nu_\mu\to \nu_\tau$, $\nu_\mu \not\to \nu_e, \nu_s$ \\
    &  &  & Regions allowed: \\
    &  &  & $\Delta{m}^2_{\rm atm} = 3.2(1.5-5)\times 10^{-3}$ eV$^2$\\
    &  &  & $\sin^2(2\theta_{\rm atm}) = 1 (>0.88)$ \\
    &  &  & Expected future data from:   \\
    &  &  & Super-K, SNO                 \\   \hline
Accelerator & $\nu_\mu,\bar{\nu}_\mu$  &  $\nu_e,\bar{\nu}_e$ appear.
          & Region allowed by LSND:\\
SBL      & &
          & $\Delta{m}^2_{\rm LSND} = 0.2 - 1.0~{\rm eV^2}$ \\
    &  &  & $\sin^2(2\theta_{\rm LSND}) = 0.003 - 0.03$ \\
    &  &  & Controversial: some LSND allowed  \\
    &  &  & regions ruled out by KARMEN \& Bugey \\
    &  &  & Expect future data from: \\ 
    &  &  & KARMEN, Mini-BooNE \\ \hline
Accelerator  & $\nu_\mu,\bar{\nu}_\mu$  & $\nu_\mu,\bar{\nu}_\mu$ survival
          & Preliminary results from K2K  \\
hadron beam &   & $\nu_e,\bar{\nu}_e$ appear
          & Three new exps online $\approx$ 2005 \\
    &  & $\nu_\tau,\bar\nu_\tau$ appear  & Determine $\nu$ parameters, CP(?)  
          \\ \hline
Accelerator  & $\nu_\mu,\bar{\nu}_\mu$  & $\nu_\mu,\bar{\nu}_\mu$ survival
          & New generation of exps \& accelerator \\
$\mu$-storage  & $\nu_e,\bar{\nu}_e$  & $\nu_\mu,\bar{\nu}_\mu$ appear
          & Accurate determination of $\nu$ parameters \\
    &  & $\nu_e,\bar{\nu}_e$ appear  & Search for CP/T effect \\ 
    &  & $\nu_\tau,\bar{\nu}_\tau$ appear &      \\ \hline
\end{tabular}
\caption{Summary of neutrino oscillation experiments.}
\label{tab:expsummary}
\end{center}
\end{table}

\vskip 3ex

\noindent

\subsection{Long baseline accelerator experiments}

Below we discuss briefly four accelerator based LBL experiments: the online 
K2K and the three approved experiments, MINOS, ICARUS and OPERA.
We also summarize the newly proposed LBL experiment J2K.

\vskip 3ex

{\bf K2K-the online LBL experiment}:

The K2K LBL neutrino oscillation experiment uses the meson neutrino beam 
produced at the KEK 12 GeV proton synchrotron. 
The average neutrino energy is 1.4 GeV which is below the tau
lepton production threshold.  
It is expected to have $10^{20}$ protons on target (POT)
in 5 years.   The physics goals of the K2K are: (a) to
check the $\nu_\mu$ survival probability, (b) to ensure that there is no
large $\nu_e$ appearance probability, and (c) to determine the oscillation
parameters.
The far detector is the 
Super-Kamiokande located 250 km west of KEK.  The experiment
was commenced in April 1999 and the first neutrino event was observed on June
19, 1999, formally established the feasibility of this new type of high 
energy
physics experiment.  

Results from the first year of running corresponding to
$2.29\times 10^{19}$ POT, were reported at ICHEP2000 \cite{Sakuda}.
A total of 27 $\nu_\mu$ interactions were observed inside the
central detector, while $40.3^{+4.7}_{-4.6}$ were expected from the
measurement of the front detector if $\nu_\mu$ is
not subject to oscillation. The data is consistent with $\nu_\mu$
oscillation with $\Delta{m}^2$ approximately $1\times 10^{-3}$ eV$^2$,
in agreement with the atmospheric result
of Super-K.  However the statistics is not significant enough to draw any
firm conclusion.  Nevertheless, it can be said that the data disfavor null
oscillation at the 2$\sigma$ level.  When more data are accumulated, an
oscillation analysis of the neutrino energy spectrum can be performed, 
a stronger conclusion on oscillations can be made and the favorable parameter 
region may be obtained.

\vskip 2ex

{\bf MINOS}\cite{MINOS}: 
The neutrino beam comes from the 120 GeV Fermilab proton synchrotron and
the detector is located in Soudan, Minnesota, USA.  The distance between 
the beam source and the detector is 730 km.  
A brief summary of the experiment is given below:
\begin{itemize}
{\setlength\itemsep{-0.6ex}
\item Goals: To (1) detect $\nu_\mu\to \nu_\tau$, (2) measure 
      $\Delta{m}^2_{\rm atm}$ and $\sin^2(2\theta_{\rm atm})$ to 10\%, and 
      (3) search for the $\nu_\mu\to \nu_e$ component of the oscillation.
\item Detector: 5 kt Iron-scintillation sandwich calorimeter with toroidal
      magnetic fields in the thin steel plates.
\item Expected number of $\nu_\mu$ charge current interactions: \\ 
\begin{tabular}{ccc} 
Beam regime   &   energy(\rm GeV)   &   CC events/kt-yr  \\
 high         &    12               &    30000  \\ 
 medium       &     6               &    1450   \\ 
 low          &     3               &     450   
\end{tabular}}
\end{itemize}
\vskip 2ex
%

{\bf ICARUS}\cite{ICARUS}  The neutrino beam CNGS (CERN Neutrino Beam to 
Gran Sasso) is 
derived from the CERN 450 GeV proton synchrotron and the detector is located 
in the Gran Sasso underground laboratory.  The average energy of the neutrino 
is less than 30 GeV and the distance traveled by the neutrino is 743 km. 
\begin{itemize}
{\setlength\itemsep{-0.6ex}
\item Goals: To do both LBL and atmospheric neutrino oscillation experiments.
      It can observe all three channels of neutrinos: $\nu_e$, $\nu_\mu$ and 
      $\nu_\tau$ and is optimized for the observation of $\nu_\tau$.
\item Detector: The detector is a liquid argon image detector.
      It involves new detector technology -- liquid TPC -- with modular
      structure.  The first module of 600 t
      will be installed soon for solar neutrino and
      atmospheric neutrino physics.  
\item Excellent $e$ identification for the $\nu_e$ appearance experiment.
\item $\nu_\tau$ appearance is a key physics goal. ICARUS expects 600 
      $\nu_\tau$
      in the liquid target.  For 4 years running 
      the experiment sensitivity will be increased to 
      $\Delta{m}^2_{\rm atm}=1.3\times 10^{-3}$ eV$^2$ and 
      $\sin^2(2\theta_{23})=1.2\times 10^{-2}$.}
\end{itemize}

\vskip 2ex

{\bf OPERA}\cite{OPERA}: The beam profile allows the CERN CNGS neutrino
beam to supply neutrinos to both ICARUS and OPERA which is also located in 
Gran Sasso.  
\begin{itemize}
{\setlength\itemsep{-0.6ex}
\item Physics Goal: Optimized for $\nu_\mu\to \nu_\tau$ oscillation and
      excellent electron identification for the $\nu_\mu\to \nu_e$ search.
\item Detector: Super module constructed out of individual scalable 
      modules. Emulsion cloud chamber construction: massive dense material
      (Pb/Fe) plates as target plus thin emulsion sheet to track
      the $\tau$ decay, target section followed by a muon detector to reduce
      background to a very low level. }
\end{itemize}

\vskip 2ex

{\bf HIPA to Super-K} (J2K)~\cite{J2K}: A new proposal has been made to use  
the proposed nuclear physics facility, the High Intensity Proton Accelerator 
(HIPA, formerly the JHF), to generate an intense neutrino beam with 
Super-K as the detector. HIPA, located about 60 km north-east of Tokyo, is 
a high intensity 50 GeV proton synchrotron accelerator,
with an intensity of
$1\times 10^{21}$POT/year. 
It will take 6 yeas to complete starting from 2001. 
We summarize the physics goals of J2K in the following:
\begin{itemize}
{\setlength\itemsep{-0.6ex}
\item Observe $\Delta{m}^2$ to an accuracy of $2\times 10^{-4}$ eV$^2$.
\item Assuming the dominant oscillation $\nu_\mu\to \nu_\tau$, 
      $\sin^2(2\theta_\mu)=\cos^4(\theta_{13})\sin^2(2\theta_{23})$ will be
      measured  with an accuracy of 0.04. 
\item $\nu_e$ appearance search: to measure
      $\sin^2(2\theta_e)=\sin^2(\theta_{23})\sin^2(2\theta_{13})$
      to 0.05.
\item $\nu_\tau$ appearance search: Very little hope unless
      $\Delta{m}^2_{32}$ is much larger than the current value of
      $3.5\times 10^{-3}$ eV$^{-3}$.
\item search for the possible presence of sterile neutrino $\nu_s$ by neutral
      current events can provide a stringent limit on the existence of
      $\nu_s$.}
\end{itemize}
Note that because of the distance, the matter 
effect is not significant and the oscillation effect is similar to that of 
the vacuum.

\newpage

\noindent

\section{Theoretical introduction to neutrino oscillation}

The experimental data available to date have provided the information needed 
for the construction of a generic framework for massive neutrinos.  Let us
first summarize the relevant experimental data before we embark on describing
the possible theoretical scenarios that embody the data.  If we accept all
data as discussed in the proceeding section we see that there are three
distinctive mass scales provided by the three categories of experiments:
the LSND, atmospheric, and solar.  The mass square 
difference (MSD) and the mixing angle in each category of experiments are 
given in Table~\ref{tab:3nudatta}.  We list the current best values which 
are mostly 
from the Super-K collaboration for the atmospheric and solar neutrinos and
the LSND collaboration for the SBL experiments.  Coming from a single 
experiment, the LSND data have to be considered as tentative and require 
confirmation. 

\vskip 2ex
\begin{table}[hbtp!]
\begin{center}
\begin{tabular}{|l|l|l|}\hline
Category of Exp  &  MSD ($|\Delta{m}^2|~({\rm eV^2}$)) 
                 &  mixing angle $(\sin^2(2\theta))$ 
       \\ \hline
LSND      &  0.2-1    & 0.003-0.03  \\ \hline
Atmospheric  &  $(1.5-5)\times 10^{-3}$ & $ >0.88$ \\ \hline
Solar  LMA & $2\times 10^{-5}$ & $\approx 0.8$ \\ \cline{2-3}             
\hspace{6ex} SMA & $5\times 10^{-5}$ & 0.001-0.01 \\ \cline{2-3}
\hspace{6ex} LOW & $8\times 10^{-8}$ & $\approx 1$ \\ \cline{2-3}
\hspace{6ex} VAC & $8\times 10^{-11}$ & $\approx 0.8$ \\ \hline
\end{tabular}
\caption{Summary of mixing parameters of the three categories of neutrino 
         oscillation experiments}
\label{tab:3nudatta}
\end{center}
\end{table}
\vskip 2ex

The striking feature of the MSD pattern given in table~\ref{tab:3nudatta} 
is that the three
category of experiments provide three well-separated mass scales.  This 
hierarchical structure of neutrino masses is similar to that of the quarks.  
It requires four distinct masses and hence the existence of at least four 
neutrino flavors.  Therefore, accepting the LSND data immediately implies  
a non-trivial extension of the neutrino sector of the SM.  The additional 
neutrino not contained in the SM is devoid of interactions with the SM
electroweak gauge bosons, usually referred to as the sterile neutrino and 
denoted as $\nu_s$.  If the LSND
data are excluded, the three SM neutrino flavors are sufficient and no 
extension of the number of neutrinos is necessary.  The latter case gives 
rise to the 3-neutrino scenario and the former the 4-neutrino scenario.  
Within either scenarios there are several cases which are different from 
one another by the ordering of the masses of the individual neutrino 
mass eigenstates.  We will illustrate the different mass assignments 
below. 

In view of the uncertainty of the LSND data, the discussion will be focused 
on the 3-flavor scenario. Comments on the 4-neutrino scenario will be made 
at the end of this section.  For a detailed introductory theoretical review
we refer to Ref.~\cite{lecture}. 

We define the neutrino states and masses by:  
\vskip 1ex
\indent\hspace{4ex} Flavor eigenstates: 
      $\nu_\alpha$, $\nu_\beta$, $\nu_\gamma$... \\
\indent\hspace{3.6ex} Mass eigenstates:
      \hspace{1ex} $\nu_1$, $\nu_2$, $\nu_3$... \\
\indent\hspace{4ex} Masses:\hspace{11ex} $m_1$, $m_2$, $m_3$...\\ 

The column vector of the flavor states will be denoted as $[\nu_\alpha]$, 
and the column of mass eigenstates by $[\nu_j]$. The two sets of states are 
related by a unitary matrix U:
\begin{eqnarray} 
[\nu_\alpha] &=& U [\nu_j] \nonumber \\
U & \equiv & \left(U_{\alpha j} \right)  
\end{eqnarray}

\begin{figure}[htb]
\begin{center}
\vspace*{-20ex}
\hspace*{-2cm}
\epsfxsize=18cm\epsfbox{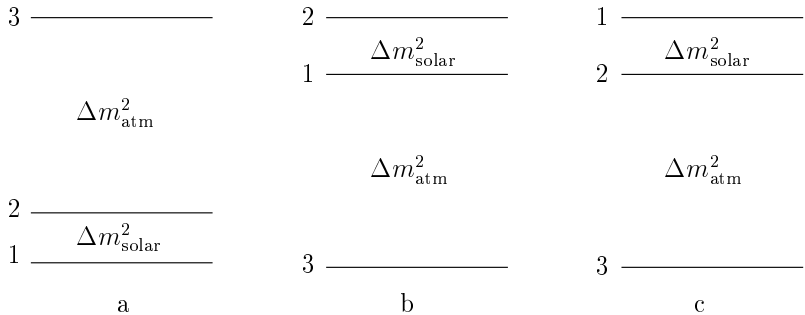}
\vspace*{-17.0cm}
\caption[] {Three possible level structures of the 3-neutrino scenario}
\end{center}
\label{fig:level3b}
\end{figure}

To illustrate the different cases of the mass assignment let us consider the 
3-flavor scenario.  
Because of the order of magnitude difference in their MSD, the solar 
and atmospheric data imply that two of the mass eigenstates lie closely in 
their mass values which we will assumed to be $\nu_1$ and $\nu_2$, with the 
third, assumed to be $\nu_3$, relatively far separated from the first two.  
Since the existing data give only the values of the MSD's not their signs, 
the mass eigenstates cannot be unambiguously identified.  Let us denote the 
mass order $m^2_j < m^2_k < m^2_l$ as the jkl-case, then there are four 
possible cases of the mass orders: 123, 213, 312 and 321.  These four 
possibilities corresponding to four possible sign assignment to 
$\Delta{m}^2_{21}\equiv m^2_2 - m^2_1$ and 
$\Delta{m}^2_{32}\equiv m^2_3 - m^2_2$.  
Future experiments will have to find out which case is correct.  Three
of the four possible level assignments,123, 312 and 321 are shown in 
Fig.~\ref{fig:level3b}


\vskip 2ex
\noindent
\subsection{Oscillation in vacuum} 

The formulae given in this subsection are valid for an arbitrary number of 
neutrino states. 

\subsubsection{Oscillation probabilities}

The oscillation probabilities are functions of the mixing matrix elements 
$U_{\alpha j}$, the MSD's $\Delta{m}^2_{kj}\equiv m^2_k - m^2_j$ with 
$m^2_k > m^2_j$, the neutrino energy $E_\nu$, and the oscillation length 
$L$. The oscillation probability of $\nu_\alpha\to \nu_\beta$ is given by 
\begin{eqnarray}
P_{\nu_\alpha\to \nu_\beta} &= & \delta_{\alpha\beta} 
        - 4\sum_{j<k} Y^{jk}_{\alpha\beta} \sin^2(\Delta_{kj})
        - 2\sum_{j<k} J^{jk}_{\alpha\beta} \sin(2\Delta_{kj})
\end{eqnarray}
where
\begin{eqnarray}
Y^{jk}_{\alpha\beta} &\equiv& Re\left( U_{\alpha j}U^*_{\alpha k}
                      U^*_{\beta j}U_{\beta k} \right) \nonumber \\
J^{jk}_{\alpha\beta} &\equiv& Im\left( U_{\alpha j}U^*_{\alpha k}
                      U^*_{\beta j}U_{\beta k} \right) \\ 
\Delta_{kj}  &=& 1.267\Delta m^2_{kj}({\rm eV}^2)
          {L({\rm km})\over E_\nu({\rm GeV})}. \nonumber 
\end{eqnarray}
$Y^{jk}_{\alpha\beta}$ is symmetric in $jk$ and $\alpha\beta$,
$J^{jk}_{\alpha\beta}$ is anti-symmetric in $jk$ and $\alpha\beta$:
\begin{eqnarray} 
Y^{jk}_{\alpha\beta} &=& Y^{kj}_{\beta\alpha} = Y^{jk}_{\beta\alpha}
                      = Y^{kj}_{\alpha\beta}  \nonumber  \\
J^{jk}_{\alpha\beta} &=& J^{kj}_{\beta\alpha} = - J^{jk}_{\beta\alpha}
                      = - J^{kj}_{\alpha\beta}
\end{eqnarray}
$CP$ and $T$ violation effects will be exhibited if
$J^{jk}_{\alpha\beta} \neq 0$.

The corresponding anti-neutrino oscillation probability can be
obtained from the above expression by the replacement $U\to U^*$, giving
\begin{eqnarray}
P_{\bar\nu_\alpha\to \bar\nu_\beta} &= & \delta_{\alpha\beta} 
        - 4\sum_{j<k} Y^{jk}_{\alpha\beta} \sin^2(\Delta_{kj})
        + 2\sum_{j<k} J^{jk}_{\alpha\beta} \sin(2\Delta_{kj})
\end{eqnarray}
This replacement for the anti-neutrinos together with the symmetry relations
of $Y^{jk}_{\alpha\beta}$ and $J^{jk}_{\alpha\beta}$ given above allows us to 
write down all the probability formulae for a given pair of neutrinos, 
$\nu_\alpha$ and $\nu_\beta$. 
\vskip 2ex

\noindent 
\subsubsection{CP/T and CPT asymmetries}

For a given pair of neutrino flavors, the probability 
$P_{\nu_\alpha\to \nu_\beta}$ is related to three other oscillations 
probabilities by CP, T and CPT transformations:
\begin{eqnarray}
CP:~~  & P_{\nu_\alpha \to \nu_\beta} \to &
         P_{\bar{\nu}_\alpha \to \bar{\nu}_\beta} \nonumber \\
T:~~~~  & P_{\nu_\alpha \to \nu_\beta} \to &
         P_{\nu_\beta \to \nu_\alpha}  \\
CPT:  & P_{\nu_\alpha \to \nu_\beta} \to &
         P_{\bar{\nu}_\beta \to \bar{\nu}_\alpha}, \nonumber 
\end{eqnarray}

Effective measurements of symmetry violations can be made by the so-called
asymmetries as defined below.  For a given pair of flavors of neutrinos
and anti-neutrinos, $\nu_\alpha$, $\nu_\beta$, $\bar{\nu}_\alpha$ and
$\bar{\nu}_\beta$, six asymmetries can be defined for 
$\nu_\alpha\neq \nu_\beta$:
\vskip 5mm
\noindent
{\bf CP asymmetries}:

\begin{equation}
A^{CP}_{\alpha\beta} = {P_{\nu_\alpha\to\nu_\beta}
              -P_{\bar{\nu}_\alpha\to\bar{\nu}_\beta} \over
              P_{\nu_\alpha\to\nu_\beta}
              + P_{\bar{\nu}_\alpha\to\bar{\nu}_\beta}}
           = {\sum_{j<k} J^{jk}_{\alpha\beta}\sin(2\Delta_{kj})\over
              2\sum_{j<k}Y^{jk}_{\alpha\beta}\sin^2(\Delta_{jk})},
\end{equation}

\vskip 5mm

\noindent
{\bf T asymmetries}

\begin{equation}
A^{T}_{\alpha\beta} = {P_{\nu_\alpha\to\nu_\beta} 
            -P_{\nu_\beta\to\nu_\alpha}\over
             P_{\nu_\alpha\to\nu_\beta} +P_{\nu_\beta\to\nu_\alpha}} 
\end{equation}

\vskip 5mm
\noindent
{\bf CPT asymmetry}

\begin{equation}
A^{CPT}_{\alpha\beta} = {P_{\nu_\alpha\to\nu_\beta}
             -P_{\bar{\nu}_\beta\to\bar{\nu}_\alpha}\over
             P_{\nu_\alpha\to\nu_\beta}
             +P_{\bar{\nu}_\beta\to\bar{\nu}_\alpha}}
\end{equation}

Three more asymmetries, one each corresponding to the above asymmetries, are
obtained by the interchange of $\nu_\alpha$ and $\nu_\beta$.  

The analytic expression of the oscillation probability assumes necessarily 
CPT symmetry and therefore gives vanishing CPT asymmetries. Hence 
given CPT symmetry there is only one independent asymmetry because of the
identity, 
\begin{equation}
A^{CP}_{\alpha\beta} = -A^{CP}_{\beta\alpha} = A^T_{\alpha\beta} 
               = -A^T_{\beta\alpha}~~~ {\rm with~CPT~invariance~in~vacuum}
\end{equation}
due to the symmetry properties of $Y^{jk}_{\alpha\beta}$ and 
$J^{jk}_{\alpha\beta}$.  
However, the CPT asymmetries given above offer a simple way to check the 
underline assumptions of the CPT invariance in the neutrino sector and 
should be made.  

In a neutrino factory, all six asymmetries can be measured for the $\nu_\mu$
and $\nu_e$ channels. But in the case of conventional meson-neutrino beams 
only the CP-asymmetries $A^{CP}_{\mu e}$ and $A^{CP}_{\bar\mu \bar{e}}$ are
accessible.

\vskip 3ex
\noindent

\subsection{The three-flavor scenario in vacuum}

The three-flavor scenario is the experimentally favorable scenario.
Although matter effects are always present in LBL experiments and their
inclusion
is necessary in order to extract precisely the oscillation parameters, the
consideration of the vacuum case is nevertheless useful.  It is a good 
approximation for experiments with shorter baselines.  Moreover, it 
gives relatively simple expressions for the various oscillation probabilities 
which provides a more transparent picture of the physics.  This allows us to 
look for the optimal experimental conditions to conduct a particular 
LBL experiment.  

\subsubsection{Basic formulae}

The unitary mixing matrix can be parameterized as:
\begin{eqnarray} 
U 
    & = & \left( \begin{array}{ccc}
    c_{12}c_{13}  & c_{13}s_{12}  &  \hat{s}^*_{13} \\
    -c_{23}s_{12} - c_{12}\hat{s}_{13}s_{23} &
      c_{12}c_{23} -s_{12}\hat{s}_{13}s_{23} & c_{13}s_{23} \\
    s_{12}s_{23} - c_{12}c_{23}\hat{s}_{13} &
    -c_{12}s_{23} -c_{23}s_{12}\hat{s}_{13}  & c_{13}c_{23}
    \end{array} \right)
\end{eqnarray}
where $s_{jk}=\sin(\theta_{jk})$, $c_{jk}=\cos(\theta_{jk})$, and
$\hat{s}_{jk}=\sin(\theta_{jk})e^{i\delta}$. $\theta_{jk}$ defined for 
$j<k$ is the mixing angle of mass eigenstates $\nu_j$ and $\nu_k$ and
$\delta$ is the CP phase angle.  The presence of a non-vanishing
$\delta$ will give rise to CP- and T-violation phenomena.

The CP/T-violation effect is given by the Jarlskog 
invariant \cite{Jarlskog}.  From the unitarity condition it can be shown 
that there is only one Jarlskog invariant in the three-neutrino scenario.  
Hence the CP/T-violation part of all the oscillation probabilities are the 
same except for a possible sign difference.  The Jarlskog invariant, denoted 
as $J(\delta)$ is defined by
\begin{equation}
J^{jk}_{\alpha\beta}=
      J(\delta)\sum_{\gamma,l}\epsilon_{\alpha\beta\gamma}\epsilon_{jkl}
\end{equation}
where $(\alpha\beta\gamma)$ in the $\epsilon$-symbol run over the three
neutrino flavors with $\epsilon_{e\mu\tau}\equiv +1$ and 
other permutations conventionally defined, and
\begin{eqnarray}
J(\delta) & = & {1\over 8} c_{13}\sin(2\theta_{12}) \sin(2\theta_{13})
                                     \sin(2\theta_{23})\sin(\delta).  
\end{eqnarray}
The CP-violation term can be rewritten as
\begin{equation}
-2\sum_{j<k}J_{\alpha\beta}^{jk}\sin(2\Delta_{kj}) 
     = -8J(\delta)\sin(\Delta_{21})\sin(\Delta_{32})\sin(\Delta_{31})
        \sum_\gamma \epsilon_{\alpha\beta\gamma}
\label{eq:cpvio}
\end{equation}

The explicit oscillation probabilities can be written done straightforwardly,
including those for anti-neutrinos with the replacement 
$U_{\alpha j}\rightarrow U^*_{\alpha j}$.

\subsubsection{Identification of mixing angles and MSD}

In terms of the experimental data, the mixing and MSD parameters can
be identified as follows:
\begin{eqnarray}
\Delta{m}^2_{21} &\equiv& \Delta{m}^2_{solar} \nonumber \\
\Delta{m}^2_{32} &\equiv& \Delta{m}^2_{atm} \nonumber \\
\theta_{12} &\equiv& \theta_{solar} \\
\theta_{23} &\equiv& \theta_{atm}.   \nonumber
\end{eqnarray}

The ranges of values of the above parameters are given in 
Table~\ref{tab:3nudatta} 
summarizing the experimental data given at the beginning of this section. 
The MSD $\Delta{m}^2_{31}$ is not independent,
\begin{equation}
\Delta{m}^2_{31} = \Delta{m}^2_{32} + \Delta{m}^2_{21}.
\end{equation}
The third mixing angle, $\theta_{13}$ is not precisely measured.  The
CHOOZ collaboration gives an upper bound of the angle,
\begin{equation}
\sin^2(2\theta_{13}) \leq 0.1.
\end{equation}

\subsubsection{Mass hierarchies, regimes of low and high mass scales}

From the  hierarchical structure of the MSD 
\begin{equation}
|\Delta{m}^2_{31}| \approx |\Delta{m}^2_{32}| >> |\Delta{m}^2_{21}|,
\end{equation}
we can classify LBL's according to the neutrino energy and the baseline 
length.  Let us define the MSD scale as the smallest MSD that
gives the maximal oscillation, i.e., $\Delta_{jk}\approx \pi/2$
\begin{equation}
\Delta{m}^2_{\bf osc}({\rm eV}^2) \equiv 
                      1.240{E_\nu({\rm GeV})\over L({\rm km})} \nonumber 
\end{equation} 
which suggests that there are two relevant experimental regions:  
$\Delta{m}^2_{\rm osc}\approx\Delta{m}^2_{21}$
and $\Delta{m}^2_{\rm osc}\approx\Delta{m}^2_{31}$.
\vskip 1ex

{\bf Low mass scale regime 
    $\Delta{m}^2_{\rm osc}\approx\Delta{m}^2_{21}$}

This region corresponds to the solar neutrino oscillation regime and for
low $E_\nu$ and long $L$, where 
$\sin^2(\Delta{m}_{31})$ and $\sin^2(\Delta_{32})$ oscillate rapidly and
can be replaced by ${1\over 2}$. Then the neutrino and anti-neutrino
oscillation probabilities can be simplified as
\begin{eqnarray}
P_{\nu_\alpha\to \nu_\beta} &=&
    \delta_{\alpha\beta}\left( 1-2|U_{\alpha 3}|^2 \right)
    +2|U_{\alpha 3}|^2|U_{\beta 3}|^2 
    -4Y^{12}_{\alpha\beta}\sin^2(\Delta_{21}) 
    -2J^{12}_{\alpha\beta}\sin(2\Delta_{21}) \nonumber \\
P_{\bar{\nu}_\alpha\to \bar{\nu}_\beta} &=&
    \delta_{\alpha\beta}\left( 1-2|U_{\alpha 3}|^2 \right)
    +2|U_{\alpha 3}|^2|U_{\beta 3}|^2  
    -4Y^{12}_{\alpha\beta}\sin^2(\Delta_{21}) 
    +2J^{12}_{\alpha\beta}\sin(2\Delta_{21})
\end{eqnarray}
In this mass scale regime, unless the baseline is very long, the neutrino 
energy will have to be low.  The experimentally interesting measurements are
the $\nu_e$ survival probability as in the solar neutrino oscillation 
experiment, and the $\nu_\mu\to \nu_e$ appearance probability:
\begin{eqnarray}
P_{\nu_e\to \nu_e} 
   & = & 1 - {1\over 2}\sin^2(2\theta_{13}) 
           - c^4_{13}\sin^2(2\theta_{12})\sin^2(\Delta_{21}),
     \nonumber \\
P_{\nu_\mu\to \nu_e}
   & = & {1\over 2}s^2_{23}\sin^2(2\theta_{13})
         +c^2_{13}c^2_{23}\sin^2(2\theta_{12})
         + 2J(\delta)\sin(2\Delta_{21}).
\end{eqnarray} 

The $\nu_e$ survive probability measures $\sin^2(2\theta_{13})$,
$\Delta{m}^2_{21}$ and $c^4_{13}\sin^2(2\theta_{12})$ if the ratio
$L/E_\nu$ can be varied so that the constant term and the oscillating
term can be separately measured.  

In this vacuum approximation the favorable distance to neutrino energy ratio 
for the LMA solution
of the solar data of $\Delta{m}^2_{\rm solar}=2\times 10^{-5}$ eV$^2$ 
is $L/E_\nu = 6.3\times 10^4$ km/GeV.  For a neutrino energy of 3 MeV,
which is in the average energy of the neutrino generated in a reactor, the
distance is 190 km.  As discussed in the preceding section, this is in
the range of the KamLAND reactor experiment. 

Note that the CP-violation term in the $\nu_\mu\to \nu_e$ oscillation is 
generally small as it is proportional to $\sin(2\theta_{13})$. 

\vskip 1ex

{\bf High mass scale regime 
    $\Delta{m}^2_{\rm osc}\approx\Delta{m}^2_{31}\approx\Delta{m}^2_{32} $}

This region is suitable for the terrestrial and accelerator neutrino 
oscillation experiments. In the leading approximation, the oscillation 
probabilities can be approximated by dropping the terms proportional to
$\Delta{m}^2_{21}$. We will also ignore the CP-violation term in this
approximation but will consider at the end of this discussion.  
\begin{eqnarray}
P_{\nu_\alpha\to \nu_\beta} &\approx & 
   P_{\bar{\nu}_\alpha\to \bar{\nu}_\beta} \nonumber \\
   &\approx & 
   \delta_{\alpha\beta}-4|U_{\alpha 3}|^2\left(1-|U_{\nu_\beta 3}|^2\right)
   \sin^2(\Delta_{32}).
\end{eqnarray}
In more detail we have 
\begin{eqnarray}
P_{\nu_\mu\to \nu_\mu} &\approx& P_{\hat\nu_\mu\to \hat\nu_\mu} 
   \approx 1-4c^2_{13}s^2_{23}(1-c^2_{13}s^2_{23})\sin^2(\Delta_{32}), 
    \nonumber \\
P_{\nu_e\to \nu_e} &\approx& P_{\hat\nu_e\to \hat\nu_e}
     \approx 1- \sin^2(2\theta_{13})\sin^2(\Delta_{32}), 
     \nonumber \\
P_{\nu_\mu\to \nu_e} &\approx& P_{\hat\nu_\mu\to \hat\nu_e}
     \approx s^2_{23}\sin^2(2\theta_{13})\sin^2(\Delta_{32}), \\       
P_{\nu_\mu\to \nu_\tau} &\approx& P_{\hat\nu_\mu\to \hat\nu_\tau}
     \approx c^4_{13}\sin^2(2\theta_{23})\sin^2(\Delta_{32}), \nonumber \\
P_{\nu_e\to \nu_\tau} &\approx& P_{\hat\nu_e\to \hat\nu_\tau} 
     \approx c^2_{23}\sin^2(2\theta_{13})\sin^2(\Delta_{32}). \nonumber
\end{eqnarray}

Due to the smallness of $\sin^2(2\theta_{13})$, in this experimental regime
the appearance probabilities $P_{\nu_\mu\to \nu_e}$ and 
$P_{\nu_e\to \nu_\tau}$ are small, probably at a few percent level. However, 
because the mixing angle 
$\theta_{23}$ is near maximal, the appearance probability 
$P_{\nu_\mu\to \nu_\tau}$ can vary from close to 1 to zero when $L/E_\nu$
varies and it can provide a good measurement for the product
$c^4_{13}\sin^2(2\theta_{23})$. The  survival probability for $\nu_e$
stays large.  But the survival probability for $\nu_\mu$ varies from
close to 1 to near zero when $L/E_\nu$ varies. 

Similar to the case of the low mass scale regime, the CP violation term
is proportional to $\sin(2\Delta_{21})$ and is neglected in the above
approximate expressions.  But it can be identified as: 
\begin{eqnarray}
\Delta{P}^{CP}_{\nu_\mu\to \nu_e} & = &
P_{\nu_\mu\to \nu_e}-P_{\bar{\nu}_\mu\to \bar{\nu}_e} \nonumber \\
  & \approx & +4J(\delta)\sin(2\Delta_{21})\sin^2(\Delta_{31})   
\end{eqnarray}
and the corresponding CP asymmetry,
\begin{equation}
A^{CP}_{\mu e} \approx {\sin(2\theta_{12})c_{13}c_{23}\sin(2\Delta_{21})\over
            2s_{23}\sin(2\theta_{13})}\sin\delta
\end{equation}

This indicates that in order to be able to see the CP-violation in the
$\nu_\mu\to \nu_e$ oscillation in this high mass scale regime, the
solar mass scale $\Delta{m}^2_{21}$ can not be too small and it is 
favorable to have a large $L/E_\nu$ to maximize the product
$\sin(2\Delta_{21})\sin^2(\Delta_{31})$. 

Identifying $\Delta{m}^2_{\rm osc}\approx |\Delta{m}^2_{32}| =
3.2\times 10^{-3} {\rm eV}^2$, then the effective
experimental probe for a baseline length of 2100 km gives  
%
%
$E_\nu \approx 5.4$ GeV.  This shows that the neutrino 
energy between 1 to 10 GeV from the H2B neutrino beam is in the optimal 
range for the probe of the atmospheric oscillation mass scale.

\vskip 3ex
\noindent
\subsection{Oscillation in matter}

The vacuum oscillation formulae is modified by the presence of matter 
along the path of the neutrino.  When neutrinos propagate through matter, 
the SM neutrinos can interact with the quarks in the nucleons and electrons 
in the atoms to undergo both elastic and inelastic scatterings.  The
inelastic scattering and the elastic scatter off the forward direction
will cause attenuation of the neutrino beam.  Since the cross sections
are extreme small, the attenuation are insignificant.  However, the
elastic scattering in the forward direction is a different matter.
Although it does not change the direction of the neutrinos in the beam,
it can modify the vacuum mixing angles and mass eigenvalues of the
SM neutrinos.  This matter effect is the well-known MSW effect~\cite{MSW},
which is not CP symmetric, hence the modifications to the
anti-neutrinos are different from those of the neutrinos.  The matter effect 
on the electron neutrino is different from that on the muon or the
tau neutrinos.  It should also be noted that the matter effect is
cumulative, in analogy with the propagation of light in a medium; the
effect will be manifested more clearly when the length of propagation
increases.  Hence a clear detection of the matter effect will require
a sufficiently long baseline. This intuitive conclusion is born out by
the explicit calculation on the effect of the matter, and the H2B's 
2100~km baseline can provide explicit checks of the matter effect.  

For the electron neutrino scattering through the matter, there are both 
charge and neutral current interactions.  The muon and tau neutrinos 
subject only to neutral current interactions.  For a sterile neutrino,
since it does not interact with the SM gauge bosons, its propagation
will not be affected by the presence of the matter.  There are special 
considerations to simplify the calculation of the matter effect.  We
give some more details below.  In the case of three-neutrino scenario
without a sterile, the neutral current interaction is the same for all
flavors.  The effect of the elastic forward scattering due to neutral
current is to give a common
phase to all flavors.  This has no effect on the oscillation and the
neutrino current interaction can be ignored.  However, the charge current
interaction, which contributes only to the elastic scattering of electron
neutrinos with the atomic electrons, has to be considered.  Because the
$e \nu_e$ and $e \hat\nu_e$ scatterings are different, matter effects
on neutrinos and anti-neutrinos are different.  Hence, the matter effect
has to be carefully removed before the information on CP-violation
can be extracted.  The effect of T-violation can be investigated more
readily, independent of the matter effect.

In a LBL experiment, as the neutrino traverses through the Earth, it 
encounters the Earth matter which may vary along the neutrino path.
The density dependent matter effect is commonly dealt by the Schr{\"o}dinger 
equation 
approach. The time evolution of the neutrino is equivalent to, and can be 
expressed as a differential equation in the distance of its propagation, 
\begin{eqnarray}
i{d\over dL}\nu_\alpha & = & {1\over 2E_\nu}\sum_\beta
      \left( (\sum_J m^2_j U_{\alpha j} U^*_{\beta j})
           + A\delta_{\alpha e}\delta_{\beta e} \right) \nu_\beta \\
A &=& 2E_\nu a_C   \nonumber \\
a_C & = & \sqrt{2}G_F n_e = \sqrt{2}G_F N_A Y_e \rho \nonumber  
\end{eqnarray}
where $a_C$ is due to the charge current effect, $G_F$ is the Fermi 
constant, $n_e$ the number density of the electron, $N_A$ the Avogadro's 
number, $\rho$(gm/cm$^3$) the matter density in units of gram per cm$^3$,
and $Y_e$ the average of number of electrons per nucleon.  The
matter density $n_e$ and hence $\rho$ depend on $L$.  For most of
the Earth density, $Y_e$ can be taken as 1/2.

The above equation can be integrated numerically for the solution. 
Note that as mentioned before we have dropped the neutral current effect, 
which is common to all the three SM neutrinos and can be ignored. The 
expressions for anti-neutrinos can be obtained by the replacement: 
$U\to U^*$ and $A\to -A$.

For constant matter density, analytic expressions of the mixing parameters 
and oscillation probabilities expressed in terms of those of the vacuum
quantities exist for the 
three-neutrino scenario~\cite{3numatterBW,3numatterNEW}. 
The explicit expressions for constant matter density are useful in 
estimating the magnitude of the matter and CP/T effects.  Especially 
the approximate expressions are transparent in their physical meaning. 
The rest of this subsection is devoted to analytic expressions for the 
case of constant matter density. 
\subsubsection{Approximate expressions for constant matter density for 
               high mass scale}

In the following we list some relevant approximate expressions in the 
case of constant matter density in the leading oscillation for 
$\Delta{m}^2_{\rm osc}\approx \Delta{m}^2_{31}\approx 
                    \Delta{m}^2_{32}>>\Delta{m}^2_{21}$, 
which are relevant to terrestrial neutrino oscillation experiments and
those for $L/E_\nu$ of the order of $(\Delta{m}^2_{32})^{-1}$, such as the
H2B.  Quantities that are subjected to modification by the approximated
matter effect will be marked by the superscript ``$m$''.  The approximate
expressions are simple enough that their physical meaning is clear. 

A comparison of the approximate expressions with the exact one will be 
made at the end of the next subsection.  The expressions given below, 
that set both $\Delta{m}^2_{21}$ and the CP phase $\delta$ to zero, are 
taken from Ref. \cite{3numatterBW}
\begin{eqnarray}
P^m_{\nu_\mu\to \nu_e} &\approx& s^2_{23}\sin^2(2\theta^m_{13})
      \sin^2(\Delta^m_{32})  \nonumber \\
P^m_{\nu_\mu\to\nu_\tau} &\approx& \sin^2(2\theta_{23})\left(
      \sin^2(\theta^m_{13})\sin^2(\Delta^m_{21})
     +\cos^2(\theta^m_{13})\sin^2(\Delta^m_{31}) 
     -{1\over 4}\sin^2(2\theta^m_{13})\sin^2(\Delta^m_{23})\right)
     \nonumber  \\
P^m_{\nu_\mu\to \nu_\mu} &=& 1 - P^m_{\nu_\mu\to \nu_e}
                               - P^m_{\nu_\mu\to\nu_\tau}  \\
P^m_{\nu_e\to \nu_\tau} &\approx& c^2_{23}\sin^2(2\theta^m_{13})
                                \sin^2(\Delta^m_{32}) \nonumber \\
P^m_{\nu_e\to \nu_e} &=& 1-\sin^2(2\theta^m_{13})\sin^2(\Delta^m_{32}) 
     \nonumber 
\end{eqnarray}
where
\begin{eqnarray}
\sin^2(2\theta^m_{13}) &=& {\sin^2(2\theta_{13})\over S^2} \\
S &\equiv& \sqrt{\left(\cos(2\theta_{13})- A/\Delta{m}^2_{31}\right)^2
            + \sin^2(2\theta_{13})} 
\end{eqnarray}
\begin{equation}
\Delta^m_{32}\equiv S\Delta_0 \nonumber,
\hspace{3ex} \Delta^m_{31}\equiv {1\over 2}
         \left(1+{A\over \Delta{m}^2_{32}}+S\right)\Delta_0,
\hspace{3ex} \Delta^m_{21}\equiv {1\over 2}
         \left(1+{A\over \Delta{m}^2_{32}}-S\right)\Delta^m_0 
\end{equation}
with
\begin{equation}
\Delta^m_0\equiv {\Delta{m}^2_{32}L\over 4E_\nu}
         =1.267\Delta{m}^2_{32}({\rm eV^2})
                {L({\rm km})\over E_\nu({\rm GeV})}
\end{equation}

The matter effect can be written explicitly for use in numerical simulation.  
A brief discussion of the earth density profile can be found in Sec. 4.1.3. 
\begin{eqnarray}                  
A   
  &=& 0.7634\times 10^{-4}\rho({\rm g/cm^3})Y_eE_\nu({\rm GeV}) 
\end{eqnarray}
where $\mathrm Y_e=0.5$ is the number of electron per nucleon.
Note that the matter resonance enhancement occurs at
\begin{equation}
E_\nu\approx 15({\rm GeV})\left(\Delta{m}^2_{31}({\rm eV^2})\over 
                  3.5\times 10^{-3}({\rm eV^2})\right)
\left({1.5({\rm g/cm^3})\over\rho({\rm g/cm^3})Y_e}\right)\cos(2\theta_{13})
\end{equation}

For the H2B experiment, taking $\rho \approx 3~{\rm gm/cm^3}$, then the
matter resonance occurs at $E_\nu\approx 15~{\rm GeV}$ which is in the
range of the HIPA neutrino beam.

As already stated that the expressions of the corresponding anti-neutrino 
oscillation is obtained by the replacement $A\to -A$.

Because of the smallness of $\sin^2(2\theta_{13})$ the above probability 
expressions show that the $\nu_\mu\to \nu_e$ and $\nu_e\to \nu_\tau$ 
appearance probabilities are small, similar 
to the value in the vacuum case as discussed early.  But they are enhanced 
due to the matter resonance effect. The $\nu_\mu\to \nu_\tau$ appearance 
probability as well as the $\nu_\mu$ and $\nu_e$ survival probabilities are 
large. 

\subsubsection{Exact results for constant matter density}

In the following we list the exact expressions of 3-flavor mixing for
constant matter density \cite{3numatterNEW}.  Quantities modified by the
exact matter effect will be denoted by a superscript or subscript ``$(m)$''.
We rewrite the Schr{\"o}dinger equation as 
\begin{eqnarray}
i{d[\nu^{(m)}_\alpha]\over dL} &=& H_M [\nu^{(m)}_\alpha] \nonumber \\ 
H_M &=& H_V + {1\over 2E_\nu}A I_{ee} \nonumber \\
    &=& {1\over 2E_\nu}U (\Delta{M}^2 + AU^\dagger{I_{ee}}U)U^\dagger \\
I_{ee} &=& \left( \begin{array}{ccc} 
                  1 & 0 & 0 \\ 0 & 0 & 0 \\ 0 & 0 & 0 
                  \end{array} \right)  
\end{eqnarray}
The matter
modified mass matrix, denoted by $\Omega$ can be diagonalized by a
matrix $V$,
\begin{eqnarray}
\Omega &=& \Delta{M}^2 + AU^\dagger{I_{ee}}U\dagger \nonumber \\
       &=& V M^2_{(m)} V^\dagger
\end{eqnarray}
where 
\begin{equation}
M^2_{(m)} = \left( \begin{array}{ccc} \lambda_1 & & \\
                    & \lambda_2 &  \\ & & \lambda_3 
                    \end{array} \right)
\end{equation}
contains the new mass square eigenvalues with the mass square eigenvalues
\cite{3numatterNEW} 
\begin{eqnarray}
\lambda_1 &=& {1\over 3}\left[x -\sqrt{x^2-3y}\left(z+\sqrt{3(1-z^2)}\right)
                        \right] \nonumber \\
\lambda_2 &=& {1\over 3}\left[x -\sqrt{x^2-3y}\left(z-\sqrt{3(1-z^2)}\right)
                        \right] \\
\lambda_3 &=& {1\over 3}\left[x + 3\sqrt{x^2-3y}\right] \nonumber 
\end{eqnarray}
where 
\begin{eqnarray}
x & = & \Delta{m}^2_{21}+\Delta{m}^2_{31}+A \nonumber \\
y & = & \Delta{m}^2_{21} \Delta{m}^2_{31}
        + A\left[ \Delta{m}^2_{21}(1-|U_{e2}|^2)
                 +\Delta{m}^2_{31}(1-|U_{e3}|^2)\right]\\
z & = & \cos\left({1\over 3}\cos^{-1}\left({2x^3 - 9xy 
    + 27\Delta{m}^2_{21} \Delta{m}^2_{31}|U_{e1}|^2}
      \over {2(x^2-3y)^{3/2}}\right) \right) \nonumber
\end{eqnarray}

 The elements of the diagonalization matrix $V$ are given by 
\cite{3numatterNEW}
\begin{equation}
V_{jj}={N_j\over D_j},\hspace{5ex}
V_{jk}={A\over D_k}(\lambda_j-\Delta{m}^2_{l1})U^*_{ej}U_{ek}
\end{equation}
where $j\neq k\neq l$ takes the values 1, 2 and 3 and
\begin{eqnarray}
N_j &=& (\lambda_j - \Delta{m}^2_{k1})(\lambda_j - \Delta{m}^2_{l1})
      -A\left[ (\lambda_j - \Delta{m}^2_{k1})|U_{el}|^2
               + (\lambda_j - \Delta{m}^2_{l1})|U_{ek}|^2 \right] \nonumber \\
D^2_j &=& N^2_j + A^2|U_{ej}|^2\left[ (\lambda_j  
                - \Delta{m}^2_{k1})^2|U_{el}|^2
           + (\lambda_j - \Delta{m}^2_{l1})^2|U_{ek}|^2 \right] 
\end{eqnarray}

Now the Hamiltonian with the matter effect can be rewritten in the same form 
as the vacuum case:
\begin{eqnarray}
H_M &=& {1\over 2E_\nu}U^{(m)}M^2_{(m)} U^{(m)\dagger} \nonumber \\
U^{(m)} &=& (U^{(m)}_{\alpha j}) = U V 
           = \left(\sum_k U_{\alpha k} V_{kj}\right)  \\
U^{(m)}_{\alpha j} &=& {(-1)^{j-1}\over D_j}\left[ N_j U_{\alpha j}
    + A U_{ej}\left( (\lambda_j - \Delta{m}^2_{k1})U^*_{el}U_{\alpha l} +
                     (\lambda_j - \Delta{m}^2_{l1})U^*_{ek}U_{\alpha j} 
              \right)\right]
\end{eqnarray} 

Now the matter case can be written in the same expression as the vacuum case 
with the matter modified mixing matrix $U^{(m)}$ and the corresponding MSD's,
\begin{eqnarray}
\Delta^{(m)}{m}^2_{21} &=&{2\over 3}\sqrt{x^2-3y}\sqrt{3(1-z^2)} 
            \nonumber \\
\Delta^{(m)}{m}^2_{32} &=&{1\over 3}\sqrt{x^2-3y}\left(3z-\sqrt{3(1-z^2)}
                                                 \right)  \\
\Delta^{(m)}{m}^2_{31} &=&{1\over 3}\sqrt{x^2-3y}\left(3z+\sqrt{3(1-z^2)}
                                                 \right) \nonumber
\end{eqnarray}

We define the oscillation argument for $j<k$:
\begin{equation}
\Delta^{(m)}_{kj}\equiv 1.267\Delta^{(m)}{m}^2_{kj}({\rm eV^2})
                  {L({\rm km})\over E_\nu({\rm GeV})}
\end{equation}

For the anti-neutrino, the corresponding quantities 
are obtained as in the neutrino case by the replacement:   
\begin{eqnarray}
U^{(m)}_{\alpha j} &\rightarrow& \bar{U}^{(m)}_{\alpha j} 
                   = U^{(m)*}_{\alpha j}(A\rightarrow -A), \\
\Delta^{(m)}_{kj} &\rightarrow& \bar{\Delta}^{(m)}_{kj} 
                   = \Delta^{(m)}_{kj}(A\rightarrow -A)
\end{eqnarray}

The oscillation probability expressions are similar to the vacuum case,
\begin{eqnarray}
P^{(m)}_{\alpha\rightarrow\beta} &=& \delta_{\alpha\beta}
    -4\sum_{j<k}Re\left(U^{(m)}_{\alpha j}U^{(m)*}_{\alpha k}U^{(m)*}_{\beta j}
           U^{(m)}_{\beta k}\right) \sin^2(\Delta^{(m)}_{kj}) \nonumber \\
    & &\hspace{4.3ex} -8J^{(m)}(\delta)\sum_\gamma 
                       \epsilon_{\alpha\beta\gamma}
\prod_{j<k}\sin(\Delta^{(m)}_{kj})  \\
P^{(m)}_{\bar{\alpha}\rightarrow \bar{\beta}} &=& \delta_{\bar{\alpha}
         \bar{\beta}}
    -4\sum_{j<k}Re\left(\bar{U}^{(m)}_{\alpha j}\bar{U}^{(m)*}_{\alpha k}
  \bar{U}^{(m)*}_{\beta j}\bar{U}^{(m)}_{\beta k}\right)
           \sin^2(\bar{\Delta}^{(m)}_{kj})  \nonumber \\
    & &\hspace{4.3ex} +8\bar{J}^{(m)}(\delta)\sum_\gamma 
                       \epsilon_{\alpha\beta\gamma}
                       \prod_{j<k}\sin(\bar{\Delta}^{(m)}_{kj})
\end{eqnarray}
where
\begin{eqnarray}
J^{(m)}(\delta) &=& \left(\Delta{m}^2_{21}\Delta{m}^2_{31}\Delta{m}^2_{32}
   \over \Delta^{(m)}{m}^2_{21}\Delta^{(m)}{m}^2_{31}\Delta^{(m)}{m}^2_{32}
                \right)J(\delta) \\ 
\bar{J}^{(m)}(\delta) &=& J^{(m)}(\delta)(A\to -A)
\end{eqnarray}
$J(\delta)$ is the vacuum expression given before.

Unlike the vacuum case, the difference $P^{(m)}_{\nu_\alpha\to \nu_\beta}$ 
- $P^{(m)}_{\bar{\nu}_\alpha\to \bar{\nu}_\beta}$ contains both the
CP-violation effect and the matter effect.  However to estimate the 
CP angle we can calculate the T-violation asymmetry in which the 
matter effect and the CP phase can be separately isolated, 
\begin{equation}
P^{(m)}_{\nu_\alpha\to \nu_\beta}- P^{(m)}_{\nu_\beta\to {\nu}_\alpha} 
  = -16J^{(m)}(\delta) \sin(\Delta^{(m)}_{21})\sin(\Delta^{(m)}_{31})
     \sin(\Delta^{(m)}_{32})\sum_\gamma \epsilon_{\alpha\beta\gamma},
\end{equation}

In the approximation of the last subsection, the T-symmetry is then given 
by
\begin{eqnarray}
A^{(m)T}_{\nu_\mu\to \nu_e} & = &
  {{P^{(m)}_{\nu_\mu\to \nu_e}-P^{(m)}_{\nu_e\to \nu_\mu}}\over
    {P^{(m)}_{\nu_\mu\to \nu_e}+P^{(m)}_{\nu_e\to \nu_\mu}}} \nonumber \\
&\approx& {8J^{(m)}(\delta)\sin(\Delta^{(m)}_{21})\sin(\Delta^{(m)}_{31})
          \over s^2_{23}\sin^2(2\theta^m_{13})\sin(\Delta^{(m)}_{32})}
\end{eqnarray}
which can be measured at a neutrino factory but not with a meson-neutrino
beam.

\vskip 1ex
Let us comment briefly on the validity of the approximate expressions 
given in the preceding subsection.  In the H2B region, i.e., $L=2100$ km 
and $E_\nu$ in the range of 1-10 GeV, the approximate expressions are  
good to within a few percent for the various oscillation probabilities.
In the energy regime of a few hundred MeV and lower, the term 
proportional to $\sin^2(\Delta_{21})$ is no longer negligible 
and it becomes eventually the dominant contribution.  Then the 
approximate expressions are no longer valid.
\vskip 2ex

\noindent
\subsection{Comment on the four-neutrino scheme}

Although the current results disfavor strongly the dominance of the 
oscillation of $\nu_e\to \nu_s$ as the mechanism for the solar neutrino 
deficit and $\nu_\mu\to \nu_s$ for the atmospheric neutrino anomaly, 
a sizable contribution of the sterile neutrino to these oscillations 
is allowed~\cite{4neutrino1}.  For a summary of the 4-neutrino fit 
of the various data, see Ref.~\cite{4neutrino}.  Because of the far 
reaching implications of such a
scenario, it is worthwhile to maintain an interest in the scenario. If
the LSND result is accepted, there are clear three different MSD's
squared mass differences of the order of magnitude: 
\vskip 2ex

$\Delta{m}^2_{\rm solar} \leq 10^{-5}~{\rm eV}^2$

$\Delta{m}^2_{\rm atm} \approx 10^{-3} {\rm eV}^2$

\begin{figure}[htbp]
\begin{center}
\vspace*{-20ex}
\hspace*{-2cm}
\epsfxsize=18cm\epsfbox{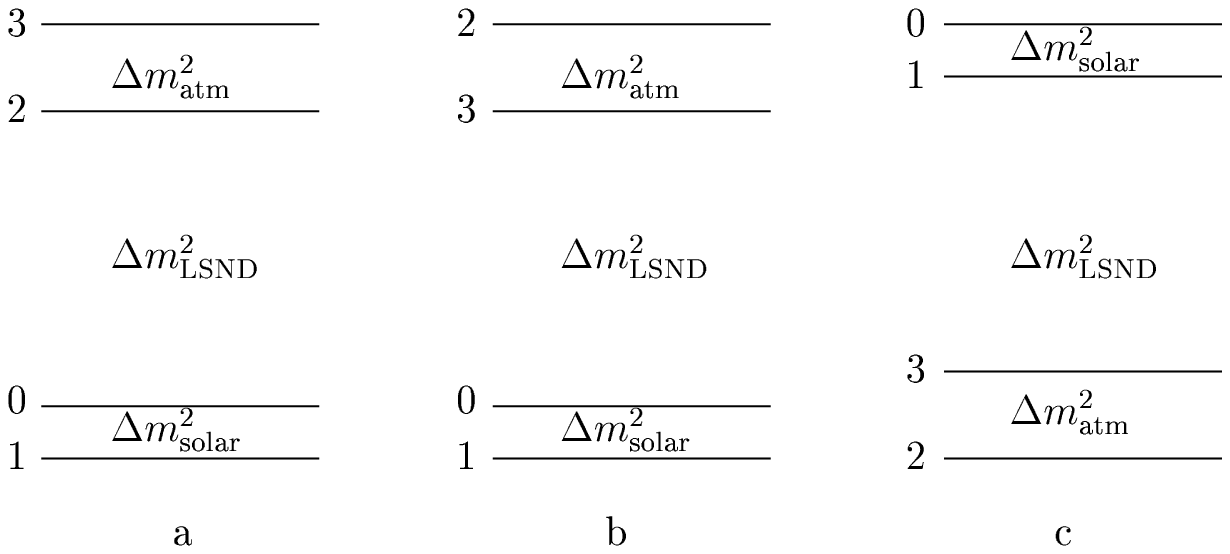}
\vspace*{-16.0cm}
\caption[] {Three possible level structures of the 4-neutrino scenario}
\label{fig:level4}
\end{center}
\end{figure}

$\Delta{m}^2_{\rm LSND} \approx 1 {\rm eV}^2$
\vskip 2ex
\noindent
which calls for at least four different neutrino masses. This has the 
implication that a right-handed or sterile neutrino exists.  The existing
data favors the so-called 2+2 scheme.  However, another possibility, i.e.,
the so-called 3+1 scheme is not completely ruled out~\cite{4neutrino1}.
In the 2+2 scheme the 4 neutrino mass
eigenstate states are divided into 2 groups each containing 2 levels.
Within each group the mass separation is relatively small in comparison
with separation between the groups. Therefore, one group gives the solar
energy scale and the other the atmospheric scale.  The scale between the
groups gives the LSND oscillation.  In the 3+1 scheme, the 4 states also
divided into two groups. One group contains 3 mass eigenstates and the other
group one state.  The 3-eigenstate group provides the solar and atmospheric
oscillation scales similar to the 3-flavor scenario.  The MSD between
the groups is the LSND scale.  Let us label the mass eigenstates as
$\nu_0$, $\nu_1$, $\nu_2$ and $\nu_3$.  Three possible level structures
for the 2+2 scheme are given in Fig.~\ref{fig:level4}


The three squared mass difference are identified as:
\vskip 2ex

$\Delta{m}^2_{\rm solar} = |\Delta{m}_{01}|$

$\Delta{m}^2_{\rm atm} = |\Delta{m}^2_{32}|$

$\Delta{m}^2_{\rm LSND}=|\Delta{m}^2_{31}|\approx |\Delta{m}^2_{21}|$, etc.

\vskip 2ex
\noindent

Again the different level structures have different signs for
$\Delta{m}^2_{kj}$, $j<k$.  In a complete determination of the neutrino
parameters, the signs of $\Delta{m}^2_{kj}$ have to be determined. 

\vskip 3ex
The four-neutrino scheme is generally treated numerically with the 
Schr{\"o}dinger equation.  In this case there are six mixing angles 
and six CP phases.  Three of the CP phases can give rise to measurable 
CP-violation effects in oscillation experiments.  For an explicit 
parameterization of the mixing matrix we refer to Ref. \cite{BDWY}.  
The matter effect is more complicated than in the three-neutrino scenario
because neutral current interactions have to be included since they are
no longer common to all neutrino flavors.  The CP effect can be sizable
and may be easier to detect.

\newpage
\noindent

\section{Fundamentals of LBL experiments and physics of H2B}

The neutrino oscillation is a system with a limited number
of degrees of freedom, it yet exhibits a multitude of interesting
phenomena. In the 3-flavor scenario, the system consists of 2 MSD, three
mixing angles and one CP phase that determine the different survival and
appearance probabilities.  The numerous neutrino experiments, solar,
atmospheric, reactor, and short baseline, are mostly looking at the
survival probabilities and often use neutrino beams that exist in nature.
In most cases there is no way to tune the neutrino beams for more desirable
experimental results.  Hence it is difficult to obtain all of the oscillation 
parameters for the entire mixing matrix, either because the statistics
are too low, or the energy is not suitable.  In long baseline experiments,
the neutrino beams are produced in the accelerator according to certain 
physics criteria so that the experiment can be conducted in a controlled
fashion.  Ideally, the distance between the neutrino source and the
detector can be chosen to maximize the physics output.  The distance can
be hundreds or thousand kilometers to allow high energy neutrino beams
to be used and still offer suitable $L/E_\nu$ values.  The LBL experiments
promise to allow for detailed analysis of the oscillation parameters so as
to provide a complete picture of the neutrino oscillations.

We will first briefly summarize some of the fundamentals of the LBL
experiment and then the possible physics that can be investigated at H2B. 
More pertinent simulation studies will be carried out with a specific detector 
later.

\vskip 2ex

\noindent
\subsection{Fundamentals of LBL experiments}

Long baseline neutrino oscillation experiments are not conventional high
energy physics experiments. Due to the extremely weak interaction cross
section of neutrinos and the long baseline between the neutrino source
and the detector, both the neutrino beam intensity and the detector mass
must be maximized in order to have the desired statistics. New technologies
for both accelerators and detectors may be required. In the following we
discuss briefly some of the fundamentals of the LBL experiment before we
start to discuss the specific neutrino beam that may be available and the
detectors that are necessary to achieve the physics goals.

\vskip 2ex
\noindent
\subsubsection{ Neutrino beams}

There are two kinds of accelerator neutrino beams: the so-called neutrino
factory from muon decays and the conventional neutrino beam from meson
decays. While meson-neutrino beams have been built by many laboratories
and the remaining technological
challenge is to increase the total power of the primary proton beam, the 
neutrino factory is a completely new concept and there are still 
a host of technical issues to be worked out.
\vskip 2ex

{\bf Neutrino factory}:  The neutrino factory delivers a neutrino beam
which contains comparable amount of $\nu_\mu$ ($\bar{\nu}_\mu$) and
$\bar{\nu}_e$ ($\nu_e$) obtained from the $\mu^-(\mu^+)$ decay in a $\mu$-storage
ring. 
\begin{equation}
\mu^-(\mu^+)\to \nu_\mu(\bar{\nu}_\mu) + \bar{\nu}_e(\nu_e)+ e^-(e^+) 
\end{equation}
Note that the presence of both the muon and electron neutrinos is not a
problem because they have opposite chiralities and therefore will not be 
a background to each other if the detector can distinguish between
positive and negative electric charges.
A description of the neutrino factory can be 
found in \cite{Geer}.

The neutrino flux at baseline L from a neutrino factory of unpolarized
muon of energy $E_\mu$ is given by
\begin{equation}
{d\Phi\over dE_\nu}= \left\{ \begin{array}{lll}
    2x^2_f (3-2x_f){n_0 \gamma^2\over \pi L^2 E_\mu} & for & \nu_\mu \\
    12x^2_f(1-x_f){n_0 \gamma^2\over \pi L^2 E_\mu} & for & \nu_e 
    \end{array} \right.
\label{eq:nuf}
\end{equation}
where $x_f = E_\nu /E_\mu$, $n_0$ is the number of useful decaying muon,
and $\gamma = E_\mu/m_\mu$ with $m_\mu$ 
being the mass of the muon.  
The average neutrino energies are given by 
\begin{eqnarray}
\langle E_{\nu_\mu}\rangle &=& 0.7E_\mu \nonumber \\
\langle E_{\nu_e}\rangle &=& 0.6E_\mu
\end{eqnarray}
Two scenarios of the number of neutrinos in a beam have been considered:
$n_0=6\times 10^{19}/$year for an entry level factory and 
$n_0=6\times 10^{20}/$year for a high performance factory.   
For a discussion of the neutrino beam spread in a 
neutrino factor together with some sample plots, see Ref. \cite{BGW}

\vskip 2ex
{\bf Meson-neutrino beam}: Neutrinos from a meson source are obtained from 
decays of pions and kaons produced by collisions of primary protons with
nuclear target.  
The secondary meson beam produced by the collision is sign selected
and then focused by a magnetic field. Mesons are then transported
to a vacuum decay pipe, whose
length depends on the desired energy of the neutrino, and finally
striking a hadron absorber further downstream.   
The primary neutrino or anti-neutrino beam consists mostly of the muon
flavor from $\pi^\pm$ and $K^\pm$ decays in the decay pipe but some
impurities of electron neutrinos are expected due to a finite branching
ratio of $\pi^\pm$, $K^\pm$ and $\mu^\pm$ decaying into electrons and
positrons plus their associated neutrinos. For example, the NuMI muon
neutrino beam at Fermilab contains 0.6\% electron neutrinos.   As for
the energy spectrum of the beam, it can be a wide-band beam covering a
broad range of energies, or narrow band beam with a selected,
well-defined energy range.   

The neutrino flux at the detector site is determined by the baseline L, 
the number of primary protons on target (POT), the proton energy $E_p$ and 
the neutrino energy $E_\nu$. The following empirical formula~\cite{malensek} 
describes the meson production from a proton beam on a nuclear target:
%
%
\begin{eqnarray}
x_M{d\sigma\over dx_M} & = & 2\pi\int x_ME_p {d^3\sigma\over dp^3}P_tdP_t 
                           \nonumber\\
                   & = & 2\pi\int {B(1-x_M)^A{1+5e^{-Dx_M}\over 
                         (1+P^2_{t}/C)^4}}P_tdP_t \nonumber\\
                   & = & (1-x_M)^A(1+5e^{-Dx_M})
\end{eqnarray}
where $E_p$ is the proton beam energy, $p$ the proton 3-momentum, $x_M$ is 
the Feynman x-variable defined as the momentum of the secondary
meson divided by the momentum of proton.  A, B, C, and D are numerical
parameters which are different for different secondary particles.  
Table~\ref{tab:mesonpara} gives their
fitted values for $\pi^+, \pi^-, K^+$, and $K^-$, taken from
Ref.\cite{malensek}.  For a wide band beam, we can take Eq.(\ref{eq:nuf}), 
i.e., the $\mathrm E_\nu^2/L^2$ behavior, to account for the  
transverse momentum spread due to pion decays. 
The neutrino flux can then be written as
\begin{eqnarray}
\Phi (E_\nu,L) \propto {E_\nu(1-x_\nu)^A(1+5e^{-x_\nu D}) 
\over L^2}
\label{eq:beam}
\end{eqnarray}
where we take $x_M \simeq x_\nu \equiv 2E_\nu/E_p$. It is interesting to note 
that this simple formula can account for the beam design of MINOS at various 
energies as discussed in Sec. 2.3.    We should remark that there is a more
complete treatment~\cite{neumesonbeam} for the energy spectrum for the 
meson-neutrino beam.  However the difference with the above express for 
$E_\nu > 1$ GeV is very small. Owing to its simpler form, we continue to use 
the above expression in the following calculations.

\begin{table}[hbtp!]
\begin{center}
\begin{tabular}{|l|c|c|c|c|}
\hline
            &  A  & B & C & D \\
\hline
$\pi^+$     & 2.4769 & 5.6817E-2 & 0.57840 & 3.0894 \\
$\pi^-$     & 3.5648 & 5.0673E-2 & 0.68725 & 5.0359 \\
$K^+$       & 1.7573 & 6.3674E-3 & 0.81771 & 5.6915 \\
$K^-$       & 5.4924 & 4.1712E-3 & 0.89038 & 2.2524 \\
\hline
\end{tabular}
\caption{Numerical parameters for meson-neutrino energy spectrum}
\label{tab:mesonpara}
\end{center}
\end{table}


\subsubsection{ Dip angle}

 To deliver the neutrino beam to the detector that is 
$L$ (direct) distance away, the beam has to  point downward at a 
dip angle $\theta_{\rm dip}$ from the horizontal plane given by 
\begin{equation}
\theta_{\rm dip} = \sin^{-1}\left(L\over 2r_{\rm E} \right),
\end{equation} 
where $r_{\rm E}=6371$ km is the average earth radius.  This is the same 
dip angle of the hadron decay pipe in the case of meson-neutrino beam.  The 
dip angle is 1.1$^\circ$ for K2K, 3.3$^\circ$ For MINOS, ICARUS and OPERA.
H2B has an oscillation length of 2143 km and the dip angle is 9.6$^\circ$.

\subsubsection{The Earth matter density profile}

\begin{figure}[htbp]
\vspace*{0.5cm}
\hspace*{2cm}
\epsfxsize=10cm \epsfbox{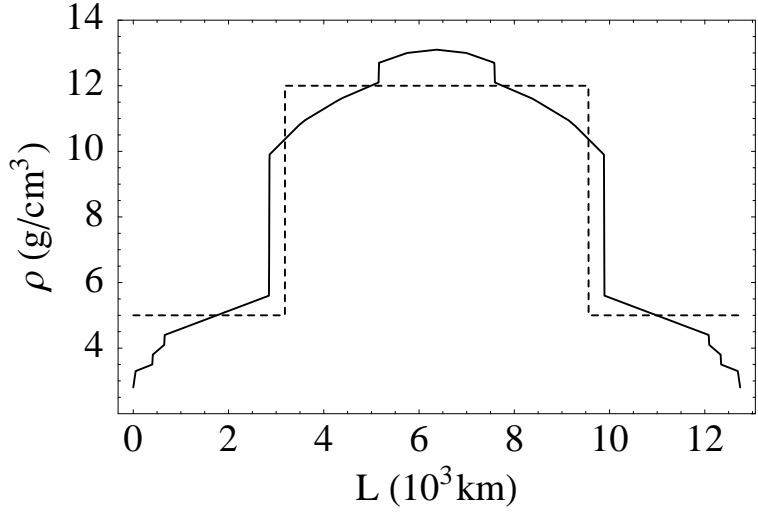}
\vspace*{0.1cm}
\caption[]{ Earth density profile.}
\label{fig:densityf}
\end{figure}

The solid earth is made of three major parts, the
crust, mantle and core.  The density increases with increasing depth $D$
from the Earth surface.  However, there are local variations of the earth 
density. 
A widely used model for the earth is the Preliminary Reference Earth Model
(PREM) provided by Dziewonski and Anderson \cite{Dziewonski}.  We list in
Table~\ref{tab:densityt} the earth densities versus the radius and the 
depth from the Earth surface, where $x\equiv r/r_E$, r is the radius at a 
given depth.  The density is in units of g/cm$^3$ and the radius and depth 
in km.  The densities given are of course the average densities. 
The extrapolation formulas are valid in the sublayers where the
mass densities vary.   
The actual densities at a given depth is given in the square bracket.
The earth density profile~\cite{Staecy} is also plotted as a function of 
the distance along a diameter from one end on the earth surface to the 
opposite end as given by the solid curve in Fig.~\ref{fig:densityf}~\cite{densityplot}. 

\begin{table}
\begin{center}
\begin{tabular}{|l|l|l|l|} \hline
region         & radius r(km) & depth D(km)    & extrapolation[density] 
                                                 (g/cm$^3$)\\ \hline
ocean              & 6371    & 0        & [1.0200] ([2.6000])  \\  
(continent)        & 6368    & 3        & [1.0200] ([2.6000])  \\ \hline
crust              & 6368    & 3        & [2.6000]  \\   
                   & 6356    & 15       & [2.6000]  \\ \cline{2-4}
                   & 6356    & 15       & [2.9000]  \\  
                   & 6346.6  & 24.4     & [2.9000]  \\ \hline\hline
LID                & 6346.6  & 24.4     & +0.6914x [3.808] \\
                   & 6291    & 80       & 2.6910 [3.3747]\\  \hline 
low velocity zone  & 6291    & 80       & +0.6914x [3.3747] \\
                   & 6151    & 220      & 2.6910 [3.3595] \\ \hline
transition zone    & 6151    & 220      & -3.8045x [3.4358] \\
                   & 5971    & 400      & 7.1089 [3.5433]\\ \cline{2-4}
                   & 5971    & 400      & -8.0298x  [3.7238]\\
                   & 5771    & 600      & 11.2494 [3.9758] \\ \cline{2-4}
                   & 5771    & 600      & -1.4836x  [3.9758]\\
                   & 5701    & 670      & 5.3197 [3.9921] \\ \hline
lower mantle       & 5701    & 670      & -6.4761x+5.6283x$^2$-3.0807x$^3$
                                          [4.3807]\\
                   & 5600    & 771      & 7.9565 [4.4432] \\ \cline{2-4}
                   & 5600    & 771      & -6.4761x+5.6283x$^2$-3.0807x$^3$
                                          [4.4432]\\
                   & 3630    & 2741     & 7.9565 [5.4915] \\ \cline{2-4}
                   & 3630    & 2741     & -6.4761x+5.5283x$^2$-3.0807x$^3$
                                          [5.4915]\\
                   & 3480    & 2891     & 7.9565 [5.5665] \\  \hline\hline
outer core         & 3480    & 2891     & -1.2638x-3.6426x$^2$-5.5281x$^3$
                                           [9.9035] \\
                   & 1221.5  & 4260.5   & 12.5815 [12.1582] \\ \hline
inner core         & 1221.5  & 4260.5   & -8.8381x$^2$ [12.764]  \\
                   & 0       & 6371     & 13.0885 [13.0885]  \\ \hline
\end{tabular}
\caption{Earth density versus the radius; $x\equiv r/r_E, D\equiv r_E -r$.}
\label{tab:densityt}
\end{center}
\end{table}
\vskip 2ex


For very long baseline experiment with high energy neutrino beams, such as
H2B, the matter effect will be important.  The deepest reach of a neutrino 
beam is given by
\begin{equation}
D=r_E(1-\sqrt{1-{L^2\over 4r^2_E}})
\end{equation}
which is 80 km in the case of H2B.  The neutrino beam will go through
mostly the earth crust but also part of the upper mantle.  On the average
the earth density along the path of H2B varies from 2.6~g/cm$^3$ at the 
initial beam entry into the earth to 3.8 g/cm$^3$ in the middle point of 
the beam and back to 2.6~g/cm$^3$ when the beam exits into the detector. 
Local variation along the beam path has to be carefully
modeled in order to investigate the CP effect.
\vskip 2ex
\noindent

\subsubsection{Interaction cross sections}

The detection of the neutrino flavor is through the charge current
interaction. For the neutrino energy which is small compared to the mass of
the W-boson, 
the charge current cross sections for the electron and muon
neutrino are given by
\begin{eqnarray}
\sigma^{(e,\mu)}_{\nu N} & = & 
         0.67\times 10^{-38}{\rm cm}^2 E_\nu(\rm GeV)  
\label{eq:crossp}
\end{eqnarray}
\begin{eqnarray}
\sigma^{(e,\mu)}_{\bar{\nu} N} & = & 
         0.34\times 10^{-38}{\rm cm}^2 E_\nu(\rm GeV)  
\label{eq:crossn}
\end{eqnarray}

\begin{figure}[htbp]
\begin{center}
\mbox{\epsfig{file=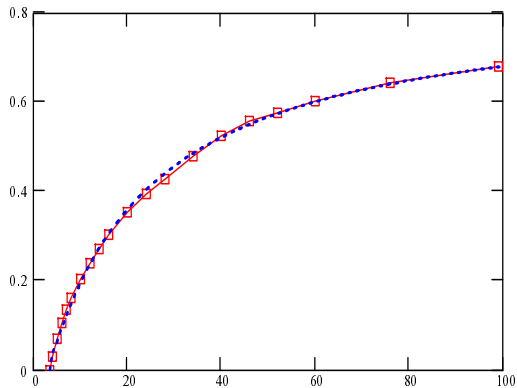, height=6cm, clip=}}
\caption[]{A fit of the numerical ratio of the $\tau$ to $\mu$ charge 
           current production cross section from the threshold to 100 
           GeV.  The boxes are the data and the dotted curve is the
           fit. The horizontal axis is the neutrino energy in GeV and the 
           vertical axis is the ratio of $\tau$ to $\mu$ production cross 
           sections. }
\label{fig:taucs}
\end{center}
\end{figure}

For the tau neutrino, the above expression is subject to a threshold 
suppression.  The threshold for the production of the tau is
\begin{equation}
E_T = m_\tau + {m_\tau^2\over 2m_N}= 3.46~{\rm GeV}
\end{equation}
The charge current production cross section of the $\tau$ is usually given 
numerically as a function of the neutrino energy.  We fit the numerical 
cross sections from the threshold to 100 GeV and 
obtained the following expression
\begin{equation}
\sigma^{(\tau)}_{\nu N}/\sigma^{(\mu)}_{\nu N} = {(E_\nu - E_T)^2\over 
                c_0 + c_1 E_\nu + c_2 E^2_\nu}\theta(E_\nu-E_T), 
\end{equation}
where $c_0=-84.988$, $c_1=18.317$, and $c_2=1.194$.
As shown in Fig.~\ref{fig:taucs}, 
the fit is good to within 3\%.  The difference occurs mostly in the 
neutrino energy region of 20-40 GeV.  The fit is valid for $E_\nu \geq 4.0$
GeV.


\subsubsection{Neutrino Statistics}

The number of neutrino events of  $E_\nu$ and flavor $\beta$, from a
neutrino beam of energy $E_\nu$ and flavor $\alpha$, to be observed
at a baseline L is given by
\begin{eqnarray}
N_{\nu_\beta} = \Phi(E_\nu,L)\sigma(E_\nu) P_{\alpha\to\beta}(E_\nu, L),
\end{eqnarray}
where $\Phi(E_\nu,L)$ is the neutrino flux including the detector size, 
$P_{\alpha\to\beta}(E_\nu, L)$ is the oscillation probability and
$\sigma(E_\nu)$ the neutrino charge current cross section. Without 
committing to an accelerator and a detector 
design we can not specify the absolute neutrino statistics
precisely.  In the 
simulation presented in the next subsection, we will take an arbitrary  
normalization. From the discussion of 
Subsection 4.1.1, we see that the relative energy spectrum of the neutrino
beam is specified.  For the neutrino factory we will not specify the parent 
muon energy, hence each neutrino energy is taken to be the average energy of 
the neutrino beam energy as also discussed in Subsection 4.1.1. Then we have 
\begin{eqnarray}
N_{\nu_\beta}   
   &\propto& \left\{ \begin{array}{ll}
   {E^3_\nu\over L^2} P_{\alpha\to\beta}(E_\nu, L) & {\rm  neutrino-factory} \\
   (1-x_\nu)^A(1+5e^{-x_\nu D}){E^2_\nu\over L^2}P_{\alpha\to\beta}(E_\nu, L)   
   & {\rm  meson-neutrino}
   \end{array} \right.
\label{eq:sta}
\end{eqnarray}
The dependence of $N_{\nu_\beta}$ on $L$ and $E_\nu$ depends on the behavior 
of the oscillation probability on these variables. The general behavior of 
oscillation probability is complicated because it is a function of two MSD's
which appear in oscillating functions. However, for LBL with neutrino energy 
of order GeV and varying within a relatively small range, the contribution 
of the solar MSD, $\Delta{m}^2_{21}$, is small and proportional to 
$(L/E_\nu)^2$.  The leading contribution comes from the atmospheric MSD 
scale, $\Delta{m}^2_{32}$.  Hence for $L/E_\nu << (\Delta{m}^2_{32})^{-1}$,
$P_{\alpha\to\beta}(E_\nu, L) \propto L^2/E^2_\nu$. 
For $L/E_\nu \approx (\Delta{m}^2_{32})^{-1}$, 
$P_{\alpha\to\beta}(E_\nu, L)$ is not subject to the $(L/E_\nu)^2$ suppression
except for the energy dependence entering the matter effect which modifies
the mixing angles.  Therefore, we have 
\begin{eqnarray}
N_{\nu_\beta} & \propto & \left\{ \begin{array}{ll}
                E  & {\rm for~} L/E_\nu << (\Delta{m}^2_{32})^{-1}  \\ 
    {E^3\over L^2} & {\rm for~} L/E_\nu \approx (\Delta{m}^2_{32})^{-1}
        \end{array} \right.   
\label{eq:nufsta}
\end{eqnarray}
for neutrino factories and
\begin{eqnarray}
N_{\nu_\beta} & \propto & \left\{ \begin{array}{ll}
   (1-x_\nu)^A(1+5e^{-x_\nu D}) 
     & {\rm for~} L/E_\nu << (\Delta{m}^2_{32})^{-1}  \\
   (1-x_\nu)^A(1+5e^{-x_\nu D}){E^2_\nu\over L^2} 
     & {\rm for~} L/E_\nu \approx (\Delta{m}^2_{32})^{-1}
     \end{array} \right.
\label{eq:pista}
\end{eqnarray}
for conventional neutrino beams by
using Eqs.(\ref{eq:nuf},\ref{eq:beam},\ref{eq:crossp},\ref{eq:crossn}). 
Naively from the above, we expect to have better results 
at higher energy for neutrino factories
while at lower energy for
conventional beams.  
In the following, we will discuss in detail where are the best place 
to do measurements
for some of the most important quantities. 
\vskip 3ex
\noindent

\subsection{ Physics of H2B}

In this subsection we will summarize in general terms the physics goals 
of H2B and compare its capability with those of different distances~\cite{okamura}.  
A preliminary study of the sensitivity of BAND on the measurements of
the physics goals, with a specific detector as an example, will be 
discussed in the next section. 
We assume that in 5 to 8 years the solar neutrino mixing parameters will 
be more accurately determined.  Also the parameters of the atmospheric 
neutrinos will be narrowed down.  A broad range of physics goals can be 
defined for H2B, depending on the neutrino beam either from a meson 
source or from a $\nu$-factory. We emphasize the advantages of an 
experiment at a very long baseline, such as 2100 km, and the neutrino energy 
in the GeV range. 
To limit the scope
of the analysis, we will only discuss the scenario of three neutrinos 
with MSW-LMA as the solution to the solar neutrino problem. 
Some of the following results will appear in Ref.~\cite{figureofmerit}.   
The earth density has been chosen to be a constant of 
$\mathrm \rho=3~g/cm^3$, other mixing parameters are set as follows:
$\Delta m^2_{32}=\Delta m^2_{31}=3\times 10^{-3} eV^2$,
$\Delta m^2_{21}=5\times 10^{-5} eV^2$,
$\theta_{12}=\theta_{23}=45^0$,
and $\theta_{13}=7^0$. The leptonic CP phase $\delta$ is set to be $90^\circ$
and the matter effect constant $A/E_\nu$ is 
$2.3\times 10^{-4} eV^2/GeV$.

\subsubsection{Mixing probability  $\sin^22\theta_{13}$}

The oscillation probability $P(\nu_{\mu}\to\nu_e)$ is a direct
measurement of $\sin^22\theta_{13}$ since
$P(\nu_{\mu}\to\nu_e)\propto \sin^22\theta_{13}$.  Of course
$\sin^22\theta_{13}$ also appears in $\nu_e\to\nu_\tau$ channel, but
experimentally it is much more difficult. The statistical significance
of  $P(\nu_{\mu}\to\nu_e)$ can be measured by the {\bf figure of merit}
defined by $P(\nu_{\mu}\rightarrow\nu_e)/\delta P(\nu_\mu\to\nu_e)$,
where the error 
$\delta P(\nu_{\mu}\to\nu_e)$ can be written as  
\begin{equation}
\delta P = \delta N_s/\Phi\sigma = \delta (N-N_b)/\Phi\sigma = 
\sqrt{\delta ^2N + \delta ^2N_b}/\Phi\sigma = 
            \sqrt{ N + r^2N^2_b}/\Phi\sigma.
\end{equation}
Here N is the total candidate event, $N_s$ the signal event and $N_b$ the 
background event. The uncertainty in estimating the background is 
represented by a parameter $\mathrm r$, 
which is typically a few percent. In this report, we  will use $r=0.1$.
If we express $N_b = f\Phi\sigma$, where $\mathrm f$ is the background 
fraction in terms of neutrino events of the original flavor, we have
\begin{eqnarray}
\delta N_s = \sqrt{P\Phi\sigma + f\Phi\sigma + r^2f^2\Phi^2\sigma^2}.
\label{eq:nserror}
\end{eqnarray}
It is clear that if $P~<<~f$, there will be no sensitivity for the given 
measurement of the signal. Generally speaking,
at shorter baseline when P is small, the effect of backgrounds is larger.
Typical values of $ f$ for water \C detectors are a few 
percent\footnote{See next section for the water \C calorimeter and 
Ref.~\cite{J2K} for water \C ring imaging detector.}.  
Contributions to $f$ from the beam are at the level of 0.6\% for 
meson-neutrino beams.  We conservatively choose 
$f=0.03$ for conventional beams and $f=0.02$ for neutrino factories.

The figure of merit $P/\delta P$ can then be written as
\begin{equation}
{P\over\delta P} = {P\over {\delta N_s / \Phi\sigma}} = 
{P\over \sqrt{ {P+f \over \Phi\sigma} + r^2f^2}}. 
\label{eq:s13}
\end{equation}
Using the analytical formula for oscillations in Sec. 3.3.2 and
Eqs.(\ref{eq:nuf}) and (\ref{eq:crossp}-\ref{eq:crossn}),
this quantity can be plotted as a function of the neutrino energy E at 
different baselines.

\begin{figure}[htbp!]
\begin{center}
\mbox{\epsfig{file=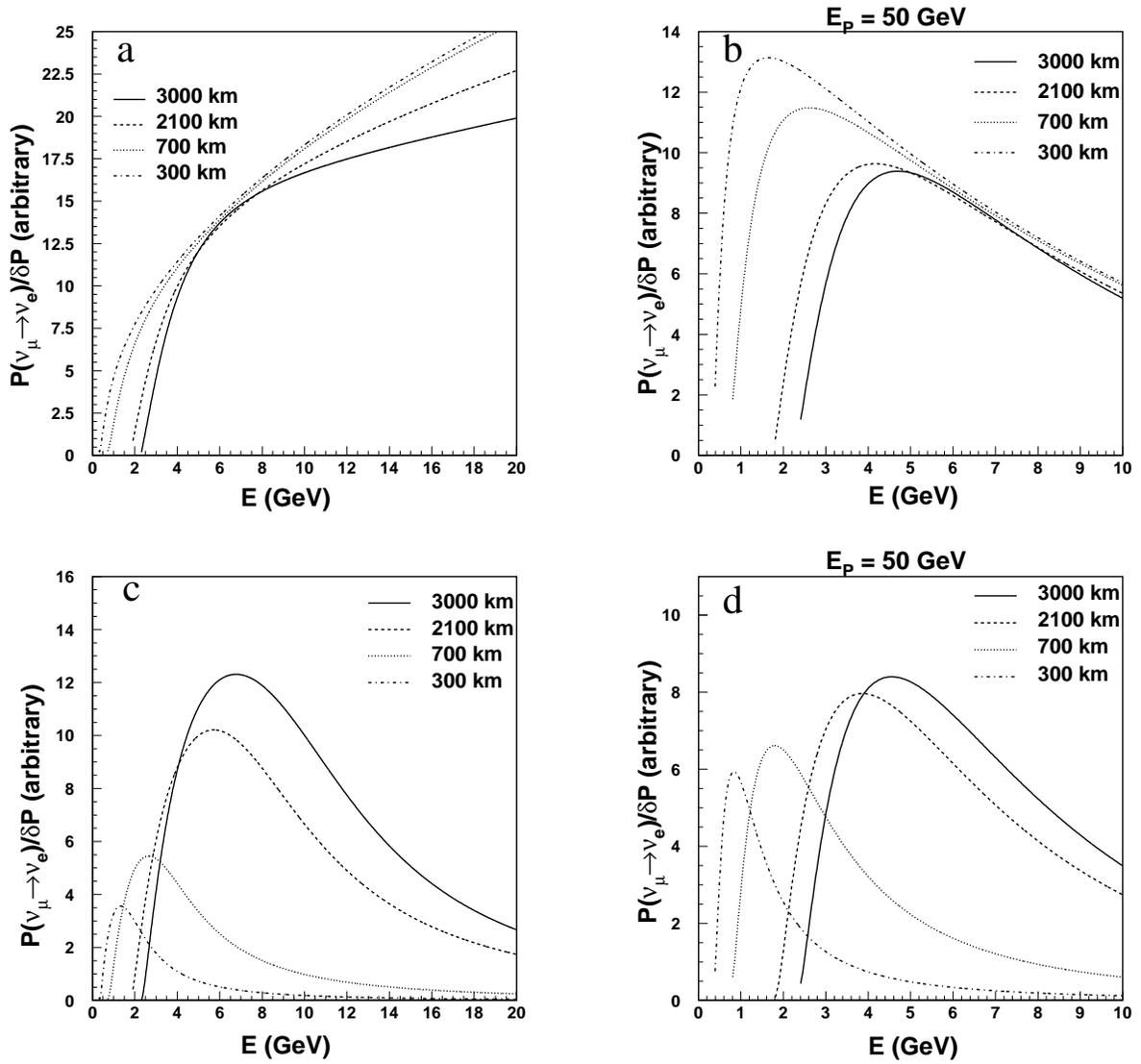,height=16cm,clip=}}
\caption{Figure of merit for $\sin^22\theta_{13}$  measurement at
         a) neutrino factories without backgrounds, b) meson-neutrino beams 
without backgrounds, c) neutrino factories with f=0.02 and r=0.1,
and d) meson-neutrino beams with f=0.03 and r=0.1.  }
\label{fig:s13}
\end{center}
\end{figure}

To see the role played by the background let us first consider the figure
of merit by ignoring the background, i.e., setting $f=0$.  The results are
given in Fig.~\ref{fig:s13}a and b. As shown in Fig.~\ref{fig:s13}a there 
is no $E_{opt}$ for a given L where the figure of merit is a maximum for 
$E_\nu \leq 20$ GeV for a neutrino factory. It would be more advantageous
to work at higher neutrino energies and shorter baselines. 
For a conventional neutrino beam, there is an $E_{opt}$ associated to each L, 
and the smaller baseline seemed to have higher figure of merit
as shown in Fig.~\ref{fig:s13}b.

The above picture is changed when the effect of the background is taken into
account. As shown in Figs.~\ref{fig:s13}c and \ref{fig:s13}d, an $E_{opt}$ 
exists for each distance with the figure of merit increasing with the 
oscillation distance for both the conventional neutrino beams and  
neutrino factories. 
For the neutrino factory, the figure of merit of the longer baselines such 
as 2100 km and 3000 km are much higher than the shorter baselines of 700 km
and 300 km.  For the meson-neutrino beam the
longer baseline is also better although the difference is not so prominent
in comparison with the case of neutrino factory.
This comparison of Figs~\ref{fig:s13}c and \ref{fig:s13}d with 
Figs~\ref{fig:s13}a and \ref{fig:s13}b demonstrates the importance of the
effect of the background in deciding the relative merits at different
baselines.

Since neutrino beams generally have a broad energy distribution, 
the correct figure of merit should integrate over the expected 
energy spectrum. While at neutrino factories, the neutrino energy
spectrum is well represented by Eq.(\ref{eq:nuf}), it is more complicated 
for meson-neutrino beams since specific beam design can alter
significantly the energy spectrum. As a general rule of thumb, 
it is always better to have a larger width of the figure of merit
curve. To this respect, it is advantageous at longer baselines as shown 
in Fig.~\ref{fig:s13}. 
A simple trial effort using Eq.(\ref{eq:nuf}) for neutrino factories
and Eq.(\ref{eq:beam}) for meson-neutrino beams shows that 
L=2100~km is preferred.

\subsubsection{Leptonic CP phase $\delta$}

As discussed in the Introduction, the leptonic CP phase for neutrinos may be 
large as the neutrino mixing angles are large.
The CP phase $\delta$ can be measured by looking at the difference of the 
oscillation probability between 
$P(\nu_\mu\rightarrow \nu_{e})$ and $P(\bar\nu_\mu\rightarrow \bar\nu_{e})$
or between
$P(\nu_e\rightarrow \nu_{\mu})$ and $P(\nu_\mu\rightarrow \nu_e)$.
While the former has to have the matter effect removed in order to obtain
the CP effect, the later, under the assumption of CPT conservation, allows
the isolation of the matter from the CP effect, but it can only be
done at a neutrino factory.

\begin{figure}[htbp!]
\begin{center}
\mbox{\epsfig{file=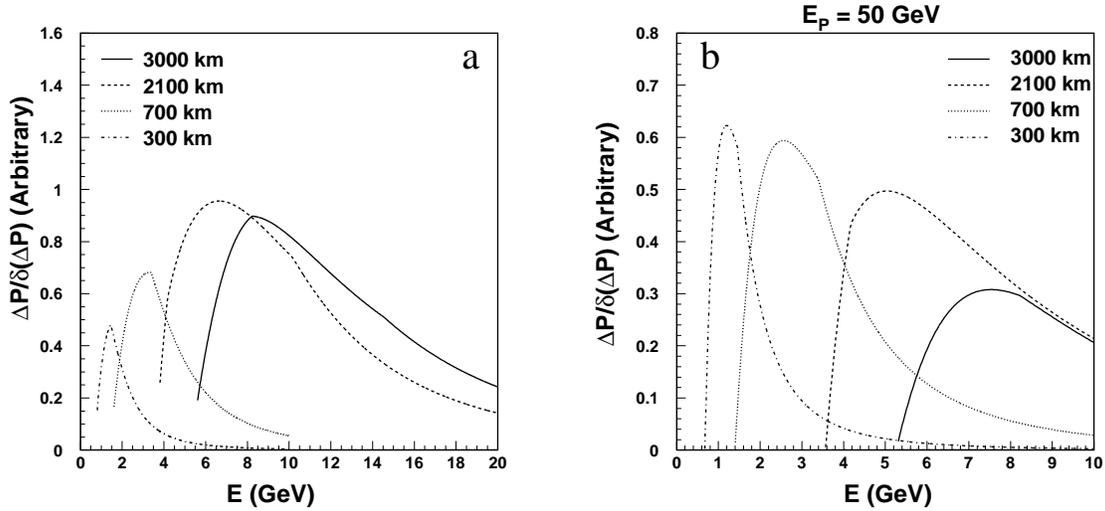, height=8cm, clip=}}
\caption{Figure of merit for CP phase measurement at a) neutrino factories 
with f=0.02 and r=0.1, 
and b) meson-neutrino beams with f=0.03 and r=0.1. }
\label{fig:cp12}
\end{center}
\end{figure}

We define 
\begin{equation}
\Delta P = P(\nu_\mu\rightarrow \nu_{e}) -  P(\bar\nu_\mu\rightarrow \bar\nu_{e})
\equiv P-\bar P
\end{equation}
and the pure CP phase can be obtained by subtracting the matter effect
$\Delta P(\delta) - \Delta P(\delta=0)$. The error of $\Delta P$ can be written as
\begin{eqnarray}
\delta (\Delta P) = \sqrt{\delta ^2P + \delta ^2\bar P} = 
                    \sqrt{{\delta ^2N_s \over\Phi^2_{\nu}\sigma^2_{\nu}} + 
                    {\delta ^2\bar N_s \over\Phi^2_{\bar\nu}\sigma^2_{\bar\nu}} }
\label{eq:perror}
\end{eqnarray}
and the figure of merit for the leptonic CP phase measurement can be written as 
\begin{eqnarray}
{\Delta P(\delta) - \Delta P(\delta=0) \over \delta (\Delta P)} 
    &=& {\Delta P(\delta) - \Delta P(\delta=0)\over \sqrt{ 
       {P+f \over\Phi_{\nu}\sigma_{\nu}} + r^2f^2 + 
       {\bar P+f \over\Phi_{\bar\nu}\sigma_{\bar\nu}} + r^2f^2}}
\label{eq:fgcp}
\end{eqnarray}
using Eq.(\ref{eq:nserror}).
Fig.~\ref{fig:cp12}a shows the figure of merit for neutrino factories with 
backgrounds as discussed above.
It is clear from the figure that, 
for every L, there is an optimum energy
for CP phase $\delta$ measurement. Once L/E is fixed at the optimum energy, 
the sensitivity to 
CP phase $\delta$ is a factor of 2 better at L=2100 km 
than that at L=300 km. 
The sensitivity to CP for conventional neutrino beams
is shown in Fig.~\ref{fig:cp12}b by using Eq.(\ref{eq:beam}) 
and (\ref{eq:crossp}-\ref{eq:crossn}). A similar optimum energy also exists, 
but the sensitivity at L=2100 km is about 20\% lower than that at L=300 km 
at their respective peak values due to unfavorable beam divergence at long 
distance. 
The integrated figure merit, however, still favors the longer baseline.

Similar conclusions can be obtained by using other quantities, such as 
$\Delta P/(P+\bar P)$, $P/\bar P$, 
$P(\nu_e\rightarrow \nu_{\mu})-P(\nu_\mu\rightarrow \nu_e)$, etc. 

>From Eq.(\ref{eq:fgcp}), we can see that the figure of merit is roughly 
proportional
to $1/\sqrt{P}$. This means that it is much better to measure the CP phase
when P is small, i.e., in the $\nu_e-\nu_\mu$ or $\nu_e-\nu_\tau$ channel,
but not the $\nu_\mu-\nu_\tau$ channel,
thanks to the small term $\sin^22\theta_{13}$ in $P(\nu_e-\nu_\mu)$ and 
$P(\nu_e-\nu_\tau)$.
In reality, we can only use the
$\nu_e-\nu_\mu$ channel due to higher experimental efficiencies.

\subsubsection{Sign of $\Delta m^2_{32}$}

Since the vacuum oscillation probability is an even function of 
$\Delta{m}^2_{kj}$, the presence of the matter effect is necessary in order 
to measure the sign of $\Delta m^2_{32}$.
The simplest way is to compare the measured probability
$P(\nu_\mu\to\nu_e)$ with expected $P^+(\nu_\mu\to\nu_e)$ and 
$P^-(\nu_\mu\to\nu_e)$ where, $P^+$ assumes $\Delta m^2_{32}>0$ and 
$P^-$ for $\Delta m^2_{32}<0$.  The channel $\bar \nu_\mu\to\bar\nu_e$ may 
be more advantageous to use if $\Delta m^2_{32} <0$.  This
measurement can only be done if $\sin^22\theta_{13}$ is sizable and 
the $\nu_e$ or $\bar\nu_e$ appearance signal is statistically significant.
Other channels, such as the $\nu_\mu\to \nu_\mu$ survival channel, 
are less sensitive due to the very small matter dependent terms. 
It has been suggested~\cite{freund} that total muons from all channels at 
neutrino factories, e.g. the $\nu_e(\bar\nu_e)\to\nu_\mu(\bar\nu_\mu)$ 
appearance and $\nu_\mu(\bar\nu_\mu)\to\nu_\mu(\bar\nu_\mu)$ survival
channels, 
be used to reduce systematic errors.  Here we confine ourselves only to 
relatively simple approaches 
to illustrate the statistical importance at different baselines.
The figure of merit in the present case can be written as 
\begin{eqnarray}
{P^+-P^- \over \delta P}= {P^+-P^- \over \delta N_s/\Phi\sigma} = 
{P^+-P^-\over \sqrt{ {P^+f\over \Phi\sigma} + r^2f^2}}.
\end{eqnarray}

\begin{figure}[htbp]
\begin{center}
\mbox{\epsfig{file=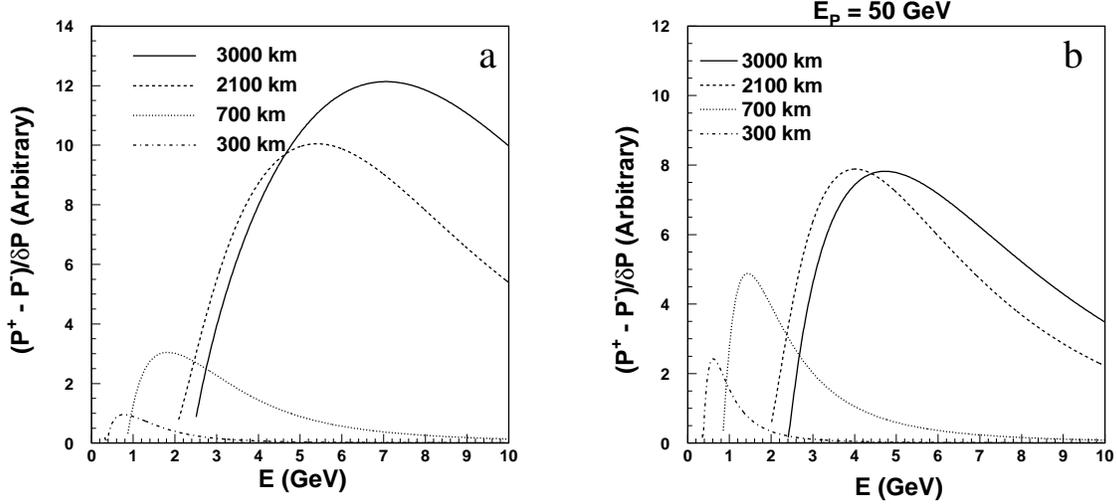,height=8cm,clip=}}
\caption{Figure of merit for the sign of $\Delta m^2_{32}$ at a) neutrino 
factories with f=0.02 and r=0.1, and
b) meson-neutrino beams with f=0.03 and r=0.1.}
\label{fig:signb}
\end{center}
\end{figure}

Fig.~\ref{fig:signb} shows this figure of merit as a function of energy at 
different 
baselines for a) neutrino factory and b) conventional neutrino beams.
It is clear from the plots that in both cases, it is much better to have an 
experiment at L=2100 than at L=300 km. 

\subsubsection{Matter effect}

The matter effect can be measured by looking at the difference between 
$\nu_\mu\to\nu_e$ and $\bar\nu_\mu\to\bar\nu_e$, although non-zero CP phase 
$\delta$ can alter the result by as much as 20\%. The survival channels 
$\nu_\mu\to\nu_\mu$ and $\bar\nu_\mu\to\bar\nu_\mu$ are not very effective for 
the present consideration since matter effects in these channels appear only 
in the non-leading terms. It is better to use the channels that involve $\nu_e$
or $\bar\nu_e$. 

\begin{figure}[htbp]
\begin{center}
\mbox{\epsfig{file=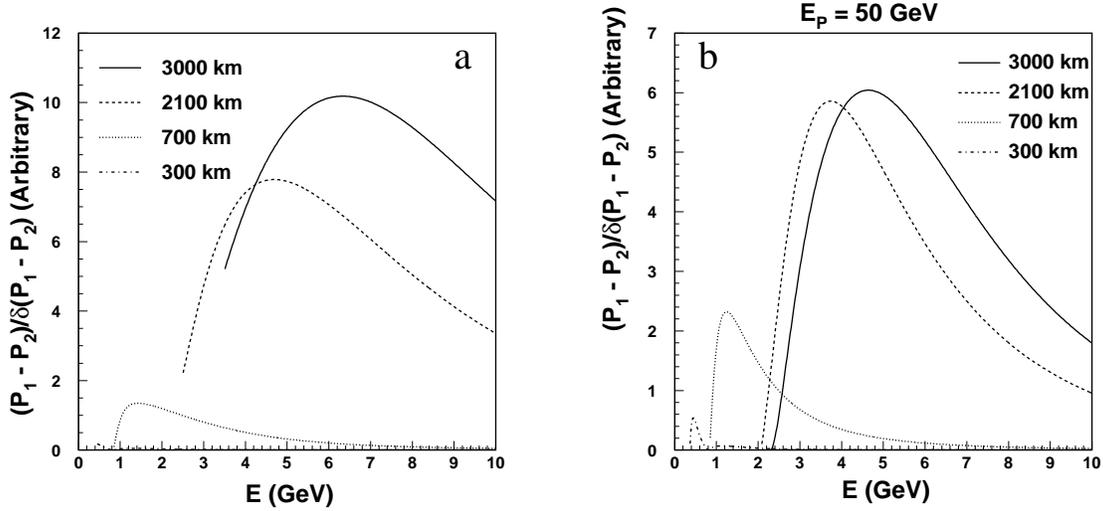,height=8cm,clip=}}
\caption{Figure of merit for the matter effect at a) neutrino factories 
with f=0.02 and r=0.1, and
b) meson-neutrino beams with f=0.03 and r=0.1. }
\label{fig:matterb}
\end{center}
\end{figure}

Denote $P_1=P(\nu_\mu\to\nu_e)$ and $P_2=P(\bar\nu_\mu\to\bar\nu_e)$,
the relevant figure of merit can be written as
\begin{eqnarray}
{P_1-P_2 \over \delta (P_1-P_2)} 
    &=& {P_1 - P_2 \over \sqrt{\delta ^2 P_1 + \delta ^2 P_2}} 
     = {P_1 - P_2 \over \sqrt{{{\delta ^2 N_{\mu} \over \Phi^2_{\nu}\sigma^2_{\nu}} 
       + {\delta ^2 N_{\bar\mu} \over \Phi^2_{\bar\nu}\sigma^2_{\bar\nu}}}}}
     =  {P_1 - P_2 \over \sqrt{{P_1+f \over \Phi_{\nu}\sigma_{\nu}} + r^2f^2+
                         {P_2+f \over \Phi_{\bar\nu}\sigma_{\bar\nu}} +r^2f^2}}.
\end{eqnarray}
Fig.~\ref{fig:matterb} shows this figure of merit as a function of energy at 
different 
baselines for a) neutrino factory and b) conventional neutrino beams.
It is clear from the plots that in both cases, it is much better to have an 
experiment at L=2100 than at L=300 km.

\subsubsection{Precision measurement of $\Delta m^2_{32}$ and
               $\sin^22\theta_{23}$}

Although $\Delta m^2_{32}$ and $\sin^22\theta_{23}$ have been measured at
super-K and hopefully will be confirmed by K2K, MINOS, OPERA and ICARUS,
it is still interesting  to measure them at different 
baselines and possibly improve the precision. 
$\Delta m^2_{32}$ and $\sin^22\theta_{23}$ can be directly related
to the survival probability $P(\nu_\mu\to\nu_\mu)$.
The figure of merit can be written as 
\begin{equation}
{1-P\over\delta P} = {1-P\over {\delta N_s / \Phi\sigma}}
                    ={1-P\over\sqrt{ {P+f\over \Phi\sigma} + r^2f^2}}.
\end{equation}

\begin{figure}[htbp]
\begin{center}
\mbox{\epsfig{file=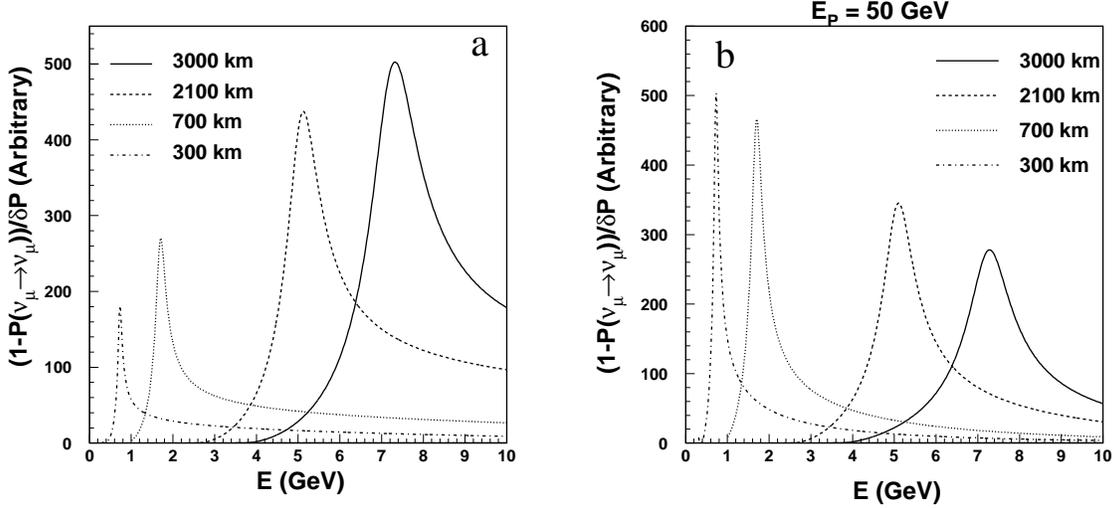,height=8cm,clip=}}
\caption{Figure of merit for $\Delta m^2_{32}$ and $\sin^22\theta_{23}$ at 
a) neutrino factories with f=0.01 and r=0.1, and b) meson-neutrino beams 
with f=0.01
and r=0.1.}
\label{fig:m23b}
\end{center}
\end{figure}

Fig.~\ref{fig:m23b} shows the figure of merit at different baseline for
a) neutrino factories and b) conventional neutrino beams. Since the
background for muon identification is generally much smaller than in the
case of the electron, we use f=0.01 for both neutrino factories and
conventional beams. A nice peak corresponding to $\Delta m^2_{32}L$ where 
the oscillation is at a maximum can be found and its position is a good 
measure of $\Delta m^2_{32}$. It is clear that for neutrino factories it 
is better at 2100 km and for conventional beams, it is better at 300 km.
The integrated figure of merit at $L=2100$ km is much higher than that
at $L=300$ km for a neutrino factory. For the meson-neutrino beam the 
longer baseline is also higher.  

\subsubsection{$\nu_\mu\to\nu_\tau$ Appearance}

The appearance of $\nu_\tau$ from a $\nu_\mu$ beam is an unambiguous proof 
of the $\nu_\mu\to\nu_\tau$ oscillation. Precision measurement of this 
oscillation probability is crucial in establishing the oscillation 
patterns and in determining whether or not $\nu_\mu\to\nu_\tau$ is the 
only dominant oscillation mode or there 
are still rooms for the $\nu_\mu\to\nu_s$ oscillation. Although we expect
that in the next few years, K2K and/or OPERA will observe the production of
$\tau$ because of the large $\nu_\mu\to \nu_\tau$ probability, it is 
desirable that BAND possesses a good $\tau$ identification capability. 
We assume that our detector can identify the $\tau$ based on 
statistics methods,  namely event selection via signature on kinematics. 
The figure of merit can be written as 
\begin{equation}
{P(\nu_\mu\to\nu_\tau)\over\delta P(\nu_\mu\to\nu_\tau)} 
                    = {P\over {\delta N_s / \Phi\sigma}}
                    = {P\over \sqrt{{P+f \over \Phi\sigma} + r^2f^2}}.
\end{equation}

\begin{figure}[htbp]
\begin{center}
\mbox{\epsfig{file=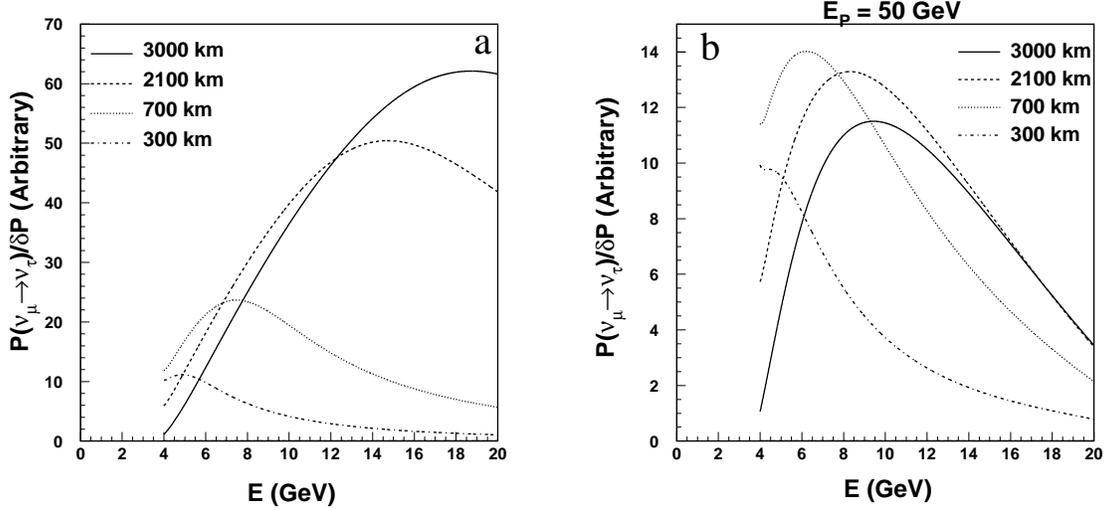,height=8cm,clip=}}
\caption{Figure of merit for tau appearance at a) neutrino factories with f=0.03 
         and r=0.1, and b) meson-neutrino beams with f=0.03 and r=0.1.}
\label{fig:mtauc}
\end{center}
\end{figure}

Fig.~\ref{fig:mtauc} shows this figure of merit as a function of energy at 
different baselines for a) neutrino factories and b) conventional neutrino
beams. Although the neutrino beam does not contribute to the background, 
it is expected to be high after event selection.
Hence we set f=0.03 for both neutrino factories and meson-neutrino beams.
Due to the threshold of tau production, the experiment has to be done at 
$\mathrm E_{\nu_\mu}>4 GeV$. That is different from what is needed 
for CP phase measurements at $\mathrm L<2000$ km. While shorter baseline 
seems better at a given energy, it is generally better to run only one 
energy for all measurements to accumulate enough statistics. To this end,
L=2100 km is much better than that at L=300 km.

\subsubsection{Exotic scenarios}

\begin{itemize}
\item {\bf Sterile neutrinos} Although the Super-K collaboration has 
demonstrated that the sterile neutrino is excluded at the 99\% CL to be the
dominant oscillation partner of $\nu_\mu$, it has not ruled out a sizable
oscillation probability for $\nu_\mu\to \nu_s$~\cite{foot}. 
The presence of $\nu_\mu\to\nu_s$ oscillation will lead to a suppressed 
survival probability $\nu_\mu\to\nu_\mu$ at $E \sim 15$ GeV 
due to matter effects. Then very long baseline experiments, such as H2B,
will have the advantages to search for the suppression. 
\item {\bf Neutrino decay} The possibility of neutrino decay as an 
explanation of Super-K's atmospheric neutrino observations has not been 
completely excluded~\cite{decay}.
However, it has been predicted that there is a significant
difference on the $\nu_\mu$ survival probability $P(\nu_\mu\to\nu_\mu)$ 
between the oscillation and decay scenarios at 
$L/E_\nu({\rm km/GeV}) \sim 400$. Therefore the figure of merit to make
the discrimination of the two scenarios is equivalent to that given in 
Fig.~\ref{fig:m23b}.  Hence we expect that K2K and MINOS will be able to
answer the neutrino decay question in the near future before we are online.
\end{itemize}

\subsubsection{Summary}

In summary, we found that by including the effect of backgrounds,
the figure of merit for the very long baseline of $L=2100$ km is higher
than that at $L=300$ km for all the measurements that we have 
considered for a neutrino factory.  For the meson-neutrino beam $L=2100$
km is still higher in most of the measurement.  
Hence according to the figures of merit, it is clearly advantageous to
do experiments at a longer baseline such as $L=2100$ km. 
In Table~\ref{tab:summ} we list all the figures of merit obtained from 
the maxima of the various curves given in 
Figs.~\ref{fig:cp12}-\ref{fig:mtauc}.

\begin{table}[hbtp!]
\begin{center}
\begin{tabular}{|l|c|c|c|c|}
\hline
            & \multicolumn{2}{c|}{neutrino factory}  & 
            \multicolumn{2}{c|}{meson-neutrino beam} \\
            & 300 km & 2100 km & 300 km & 2100 km \\
\hline
$\sin^22\theta_{13}$      & 3.5 & 10.5 & 6.0 & 8.0 \\
CP phase $\delta$        & 0.5 & 0.95 & 0.62 & 0.5 \\
sign of $\Delta m^2_{32}$& 1.0 & 10. & 2.5 & 8.0 \\
 matter effects          & 0.2 & 8. & 0.5 & 6.0 \\
$\Delta m^2_{32}$ and $\sin^22\theta_{23}$ & 180 & 450 & 500 & 350 \\ 
$\tau$ appearance        & 10  & 50  & 9.5 & 13.5 \\ 
\hline
\end{tabular}
\caption{Summary of relative figure of merits for various measurement
         at baselines L=300 and 2100 km.}
\label{tab:summ}
\end{center}
\end{table}

To put things in perspective, a few remarks are due.  A 3$\sigma$ 
statistical significance corresponds to value 3 in this table but 
unfortunately, the values of the figures of merit as given in 
Table~\ref{tab:summ} are only relative.  Eventually the absolute
statistics should be used to obtain the absolute figure of merit. 
There are two absolute normalization factors which we treat as free
parameters in the present discussion: one for neutrino factory and  
the other for conventional neutrino beams. These parameters  
depend on the beam intensity, the detector mass, efficiency, etc...  
In addition, we also need to know accurately the 
oscillation parameters such as $\sin^2(2\theta_{13})$, $\delta$, 
$\Delta m^2_{12}$, etc. However independent of these parameters it is 
clear from the table that
at neutrino factories, L=2100 km is always better than L=300 km. 
For the meson-neutrino beams, it is better at L=2100 km for 
$sin^22\theta_{13}$, the sign of $\Delta m^2_{32}$, the matter effects 
and tau appearance. For CP phase measurement, it is comparable at
all different baselines while for $\Delta{m}^2_{32}$ 
and $\sin^2(2\theta_{23})$,  L=300 km has a higher figure of merit.
However, when the integrated figures of merit are considered
the $L=2100$ is always preferred. 

\newpage
\noindent
\section{Far detector}

As we discussed before, our main physics objectives include
the measurements of $\sin\theta_{13}$, $\Delta m^2_{31}$, the leptonic
CP phase $\delta$
and the sign of $\Delta m^2_{32}$. 
All of these quantities can be obtained through the 
survival probability $\mathrm P(\nu_{\mu}\rightarrow\nu_{\mu})$ and
the appearance probability 
$\mathrm P(\nu_{\mu}(\nu_e)\rightarrow \nu_e(\nu_{\mu}))$ and
$\mathrm P(\bar\nu_{\mu}(\bar\nu_e)\rightarrow \bar\nu_e(\bar\nu_{\mu}))$.
To measure these quantities, a
detector should:
1) be able to identify charge leptons: e, $\mu$ and $\tau$;
2) have good pattern recognition capabilities for background rejection;
3) have good energy resolution for event selection and for determining 
$\mathrm P_{\alpha\rightarrow\beta}(E)$ as a function of the neutrino
energy;
4) be able to measure the charge for $\mu^{\pm}$ in the case of
 $\nu$ factories; and
5) has a large mass (100-1000~kt) at an affordable price.

\begin{table}[htbp!]
\begin{center}
\begin{tabular}{|l|c|c|c|c|}
\hline
              & Iron        & Liquid      & Water Ring & Under Water/Ice  \\
              & Calorimeter & Ar TPC      & Imaging    & \v Cerenkov counter \\ 
\hline
Mass          &  5-50 kt        & 1-10 kt     & 50-1000 kt  & 100 Mt   \\
Charge ID     &  Yes             & Yes            & ?          & No       \\
E resolution    & good             & very good    & very good  & poor    \\
Examples      & Minos            & ICARUS       & Super-K, Uno & Amanda, Icecube \\
              & Monolith         &              & Aqua-rich  & Nestor,Antares \\
Price (\$)    & 10M/kt           & 30M/kt       & 2M/kt      & 1K/kt         \\
\hline
\end{tabular}
\caption{Currently proposed detector for $\nu$ factories and conventional 
$\nu$ beams.}
\label{tab:comp}
\end{center}
\end{table}

Currently there are four types of detectors proposed\cite{nuf,dick}, as listed
in table~\ref{tab:comp}, for large LBL experiments.
As seen from the table,  they are generally expensive
and some of them are difficult to have magnet in order to measure the sign
of the electric charge of the muon. However,
much cheaper detectors, say at a price target of 0.2 M/kt,
are not impossible. They
are very attractive since they allow us to build a detector of
100 - 1000 kt without much difficulties. Here we describe one such option which was 
first proposed in ref.~\cite{wang}. Of course a significant amount
of R\& D are needed for this 
option and we are open to other detector designs.

\subsection{Water Cerenkov Calorimeter}

\begin{figure}[htbp!!]
\begin{center}
\mbox{\epsfig{file=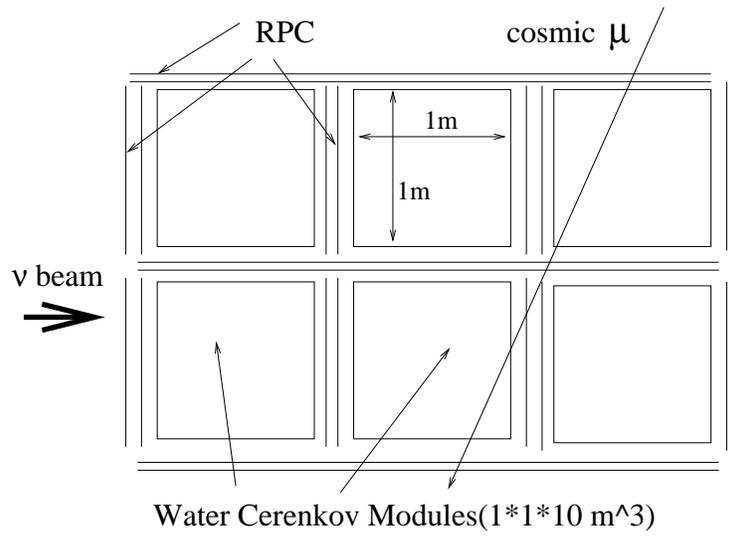,height=7.cm}}
\caption{Schematic of water \C calorimeter }
\label{fig:det}
\end{center}
\end{figure}

Water \C ring image detectors have been successfully employed 
in large scale, for obvious economic reasons,
by the IMB and the Super-Kamiokande collaborations.
However a substantial growth in size 
beyond these detectors appears problematic because of 
the cost of excavation and photon detection. 
To overcome these problems, we propose here 
a water \C calorimeter with a modular structure,
as shown in Fig.~\ref{fig:det}.

\begin{figure}[htbp!]
\begin{center}
\mbox{\epsfig{file=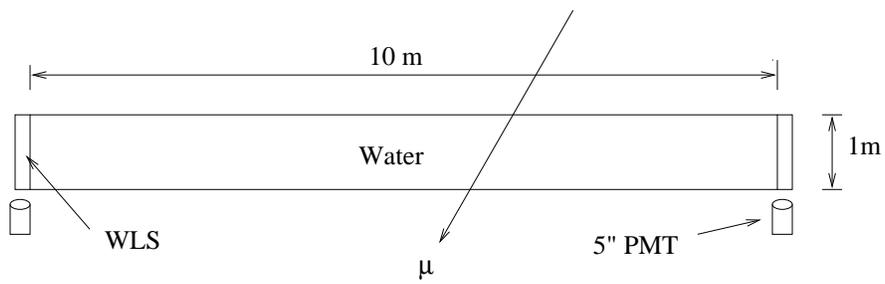,height=3.6cm}}
\caption{Schematic of a water tank.}
\label{fig:tank}
\end{center}
\end{figure}

Each tank has  
dimensions $\mathrm 1\times 1\times 10 m^3$, holding a total of 10~t of water.
The exact segmentation of water tanks is to be optimized based on the 
neutrino beam energy,
the experimental hall configuration, the cost, etc. For simplicity, we 
discuss in the following a module of 1~m in thickness,
corresponding to 2.77 X$_0$ and 1.5 $\lambda_0$. 
The water tank is made of PVC with aluminum lining.
\v Cerenkov light is reflected by the aluminum lining and transported
towards the two ends of the tank, which are covered by 
wavelength shifter(WLS) plates. 
Light from the WLS is guided to a 5'' photon-multiplier tube(PMT), as shown in 
Fig.~\ref{fig:tank}.
The modular structure of such a detector allows it to be placed 
at a shallow depth in a cavern of any shape(or possibly even at surface), 
therefore reducing the excavation cost, in particular for the presently planned
H2B site where a hall is available and a large scale excavation is unnecessary.
The photon collection area is also
reduced dramatically, making it possible to build a large detector 
at a moderate cost, which is targeted at \$0.2M/kt.

A through-going charged particle emits about 
20,000 \C photons per meter. Assuming a light attenuation length in water of
20m and a reflection coefficient of the Aluminum lining of 90\%, we obtain 
a light collection efficiency of about 20\%. 
Combined with the quantum 
efficiency of the PMT(20\%), the WLS collection efficiency(25\%) and an additional 
safety factor of 50\%, the total light collection efficiency is about 0.5\%.
This corresponds to 100 photoelectrons per meter, 
which can be translated to a resolution of $\mathrm 4.5\%/\sqrt{E}$.
This is slightly worse than the Super-Kamiokande detector and 
liquid Argon TPC but much better than
iron calorimeters\cite{nuf}.

If this detector is built for a $\nu$ factory,  
a tracking device, such as
Resistive Plate Chambers (RPC)\cite{rpc} will be needed between water tanks 
to identify the sign of charge.
RPCs can also be
helpful for pattern recognition, to determine precisely muon directions,
and to identify cosmic-muons for either veto or calibration. 
The RPC strips will run in both X- and Y-directions with a width of 4 cm.
A total of $\sim 10^5 ~m^2$ is needed for a 100 kt detector, which is more than
an order of magnitude larger than the current scale\cite{rpc}.
R\&D efforts would be needed to reduce costs.

\begin{figure}[htbp!]
\begin{center}
\mbox{\epsfig{file=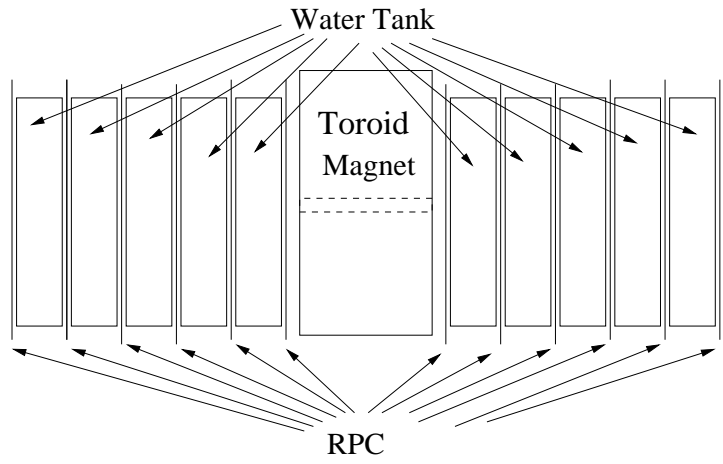,height=6.cm}}
\caption{Schematic of a toroid magnet}
\label{fig:magnet}
\end{center}
\end{figure}

The magnet system for such a detector can be segmented in order to minimize dead 
materials between water tanks.
If the desired minimum muon momentum is 5 GeV/c, the magnet must be segmented
every 20 m. Detailed magnet design still needs to be worked out; here we just 
present a preliminary idea to initiate the discussion.
A toroid magnet similar to that of Minos, as shown in Fig.~\ref{fig:magnet}, 
can produce a 
magnetic field $\mathrm B>1.5~\mathrm T$, for a current $\mathrm I> 10^4$~A.
The thickness of the magnet needed is determined by the error from
the multiple scattering:
$\mathrm \Delta P/P = 0.0136\sqrt{X/X_0}/0.3BL$,
where L is the thickness of magnet.
For L=50 cm, we obtain an error of 32\%.
The measurement error is given by 
$\mathrm \Delta P/P \simeq \delta\alpha/\alpha=
\sigma P/0.3rBL$,
where r is the track length before or after the magnet
and $\sigma$ is the pitch size of the RPC. For P=5 GeV/c, $\sigma=4$ cm and r=10 m,
the measurement error is 9\%, much smaller
than that from multiple scattering. 
It should be noted that $\mathrm P_{\mu}$ is also measured from the range. By requiring
that both $\mathrm P_{\mu}$ measurements are consistent, we can eliminate most of the 
fake wrong sign muons. 
The iron needed for such a magnet is about 20\% of the total mass of the water.

The cost of such a detector is moderate compared to other types of detectors, enabling
 us
to build a detector as large as 100 - 1000 kt. 
The combination of size, excellent energy resolution and 
pattern recognition capabilities makes this detector very attractive. 
\subsection{Performance of Water Cerenkov Calorimeter}

\begin{figure}[htbp!]
\begin{center}
\mbox{\epsfig{file=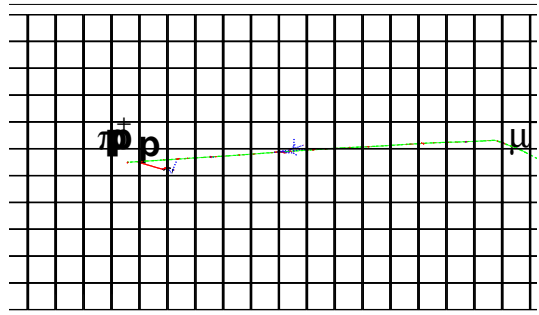,height=7.cm}}
\caption{A typical $\nu_\mu$ CC events in the detector.}
\label{fig:event}
\end{center}
\end{figure}

A full 
GEANT Monte Carlo simulation program and the Minos neutrino event 
generator have been used to study the performance of such a detector.
Fig.~\ref{fig:event} shows a typical $\nu_\mu$ charge current (CC) 
event in our detector.

\begin{figure}[htbp!]
\begin{center}
\mbox{\epsfig{file=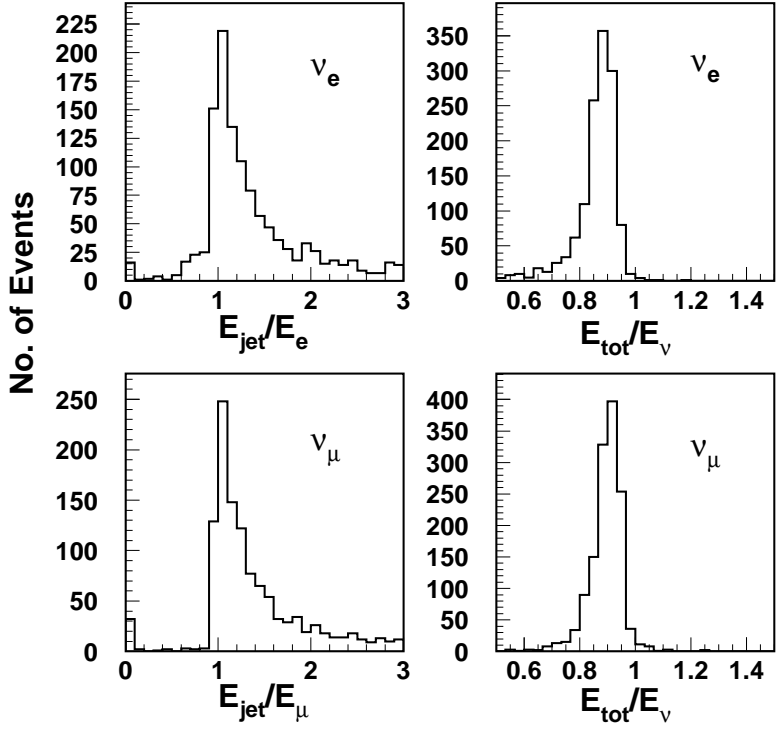,height=10.cm}}
\caption{The reconstructed jet energy and total visible energy. 
The fact that $\mathrm E_{jet}/\mathrm E_{lepton}$
peaks around one shows that the jet reconstruction algorithm 
finds the lepton from CC events. The fraction of total visible energy 
to the neutrino
energy indicates that we have an energy resolution better than 10\% for 
all neutrinos. The bias is due to invisible neutral hadrons and charged 
particles below \C thresholds.}
\label{fig:ejet}
\end{center}
\end{figure}

A CC $\nu$ signal event is identified by its 
accompanying lepton, reconstructed as a jet.
Fig.~\ref{fig:ejet} shows the jet energy normalized by the energy of the lepton.
It can be seen from the plot that leptons from CC events 
can indeed be identified and the jet reconstruction algorithm works properly.
It is also shown in the figure that the energy resolution of the neutrino CC events
is about 10\% in both cases.

\begin{figure}[htbp!]
\begin{center}
\mbox{\epsfig{file=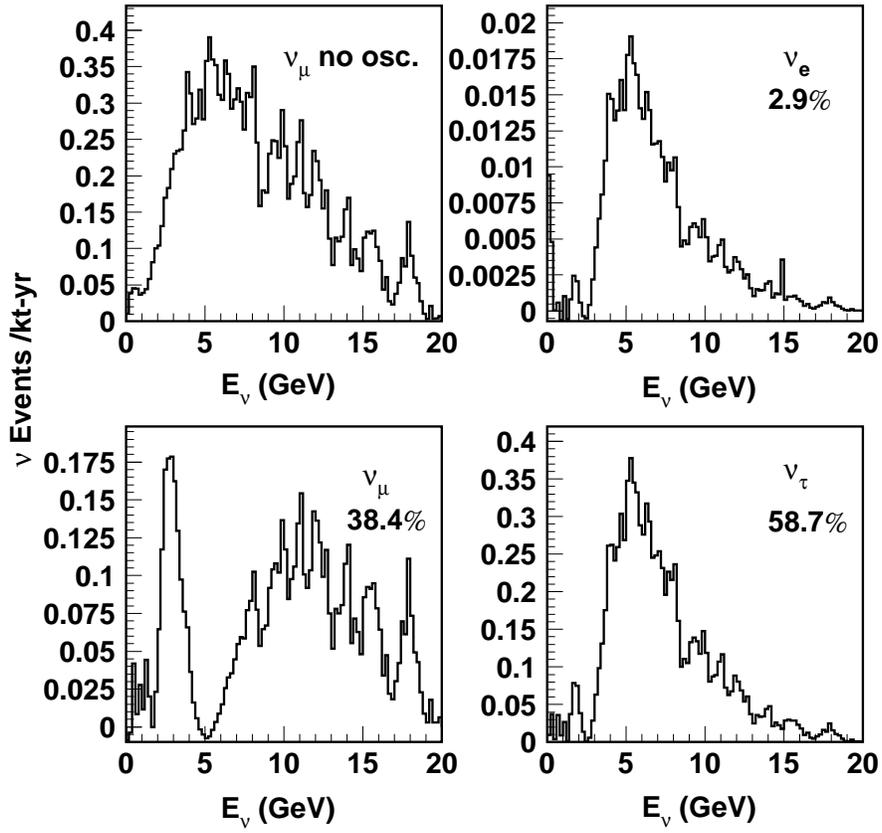,height=14.cm}}
\caption{Beam profile of HIPA-Beijing }
\label{fig:jhfbeam}
\end{center}
\end{figure}

The meson-neutrino beam profile from HIPA has been discussed in Sec. 4
and the energy spectra of visible $\nu_{\mu}$ CC events are shown in 
Fig.~\ref{fig:jhfbeam}. A total of at least 30 event/kt per year is 
expected at BAND. For the present report we will not discuss the 
detector performance at the neutrino factory.

\begin{figure}[htbp!!!]
\begin{center}
\mbox{\epsfig{file=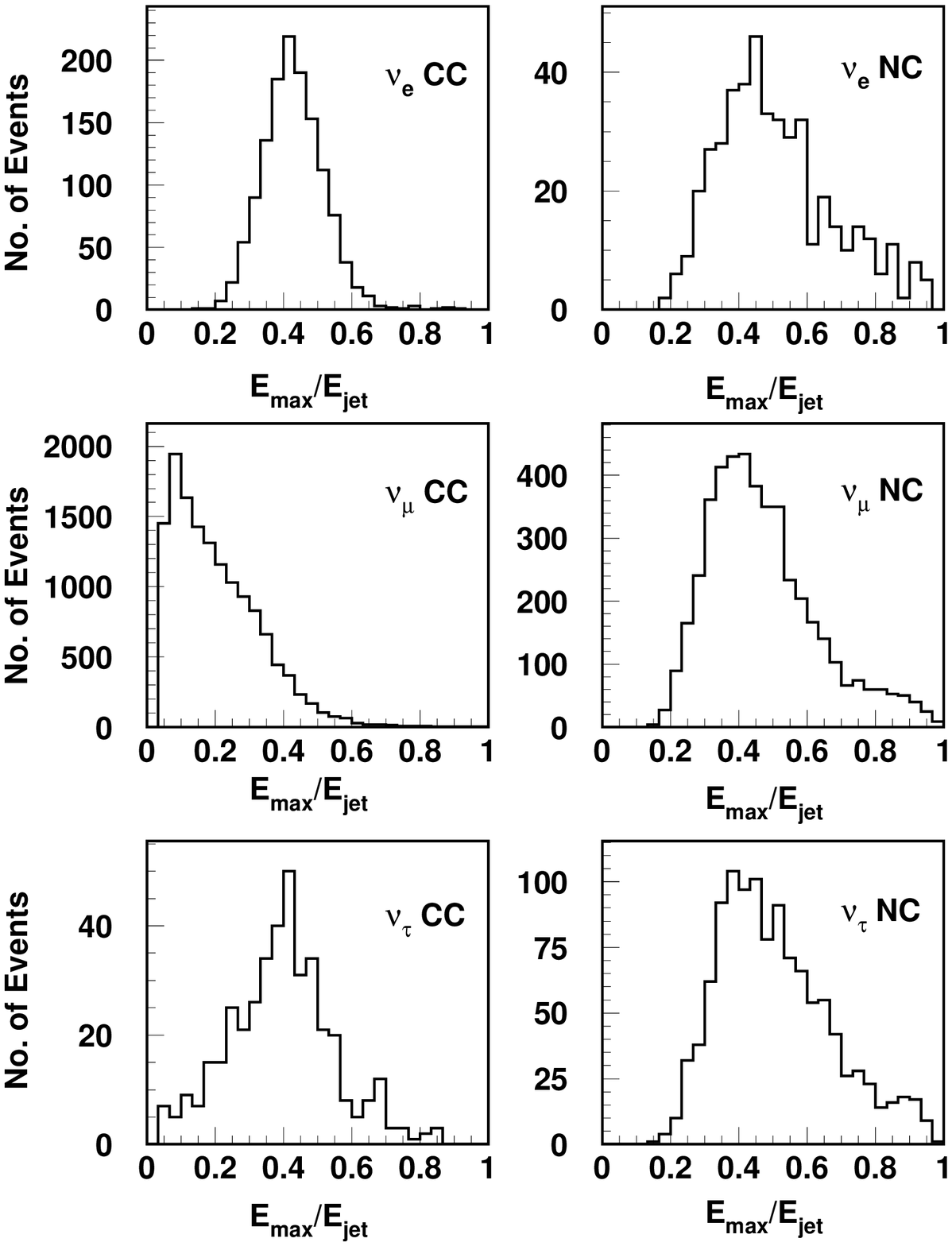,height=18.cm}}
\caption{The maximum energy deposit in a cell 
normalized to the jet energy, $\mathrm E_{max}/\mathrm E_{jet}$,
for various type of neutrino events.}
\label{fig:emax}
\end{center}
\end{figure}

\begin{figure}[htbp!!!]
\begin{center}
\mbox{\epsfig{file=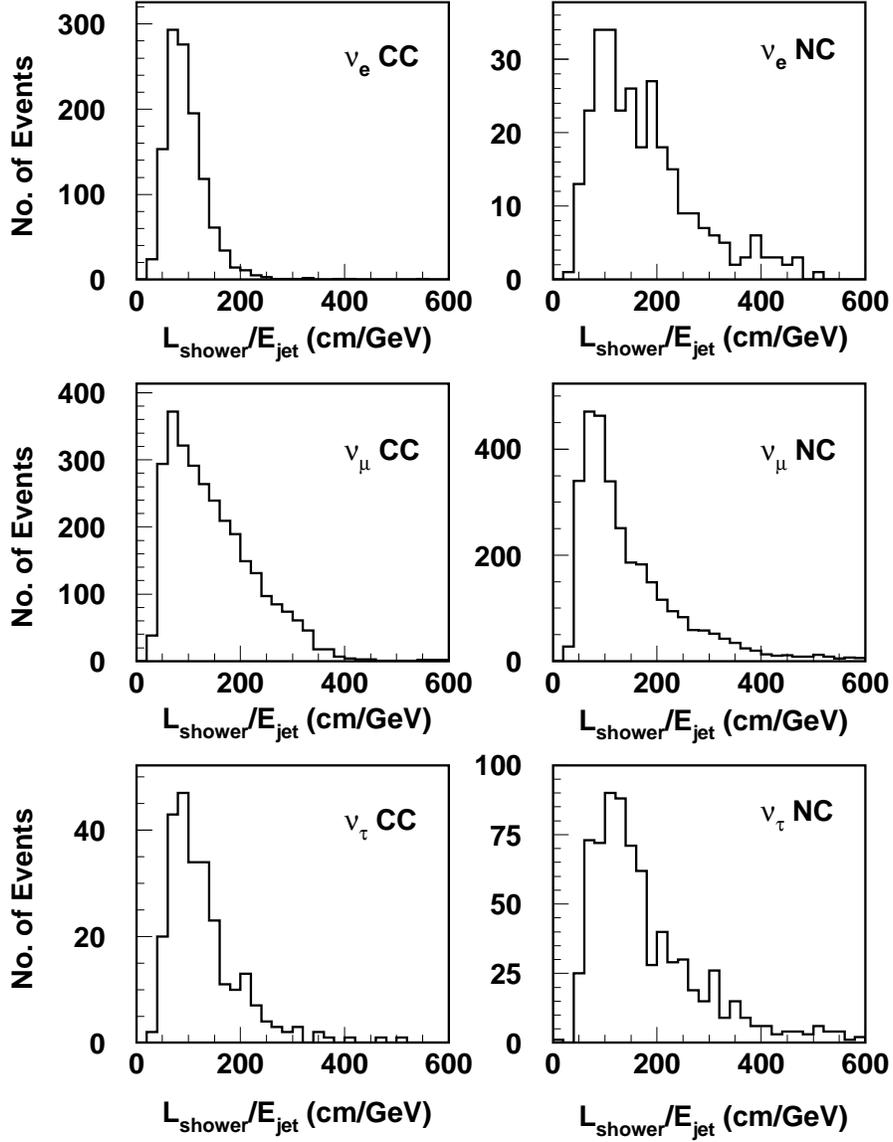,height=18.cm}}
\caption{The longitudinal length of a jet
normalized to the jet energy, $\mathrm L_{shower}/\mathrm E_{jet}$,
for various type of neutrino events.}
\label{fig:zshow}
\end{center}
\end{figure}

\begin{figure}[htbp!!!]
\begin{center}
\mbox{\epsfig{file=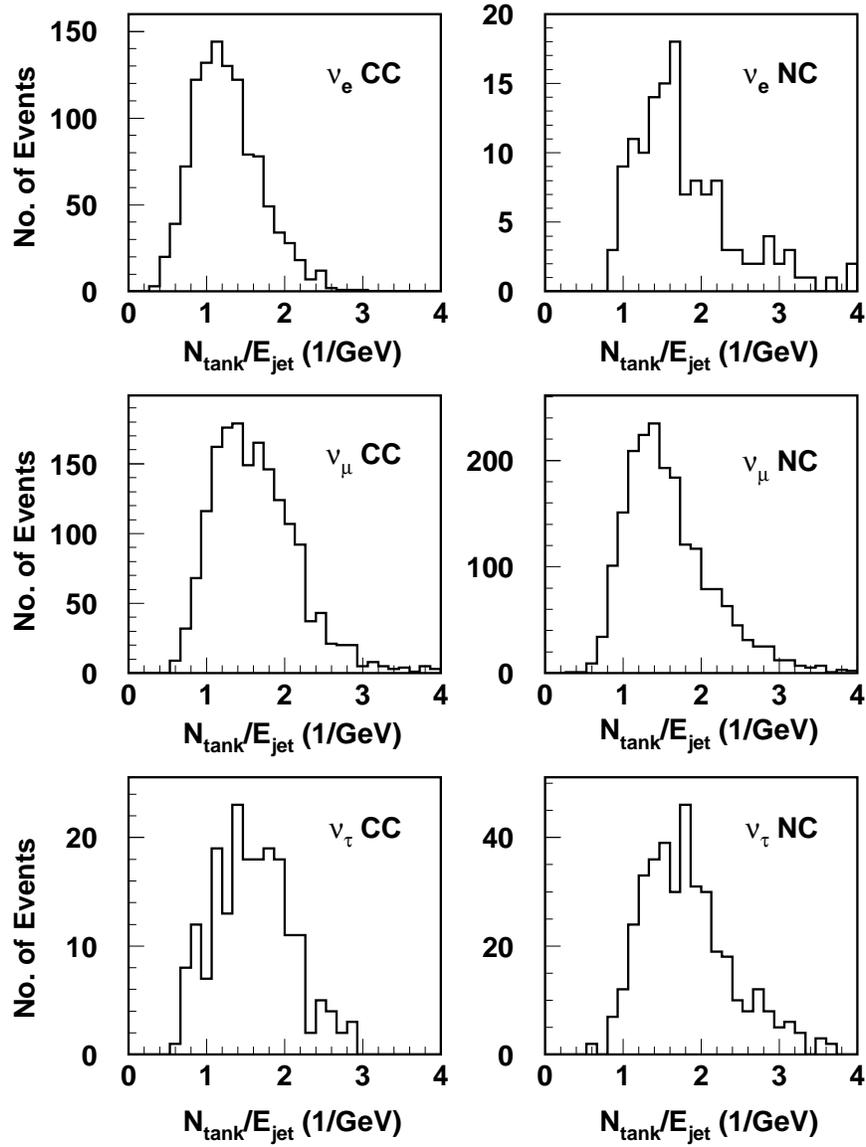,height=18.cm}}
\caption{The number of cells fired normalized to the jet energy,
$\mathrm N_{tank}/\mathrm E_{jet}$,
for various type of neutrino events}
\label{fig:ncell}
\end{center}
\end{figure}

\begin{figure}[htbp!!!]
\begin{center}
\mbox{\epsfig{file=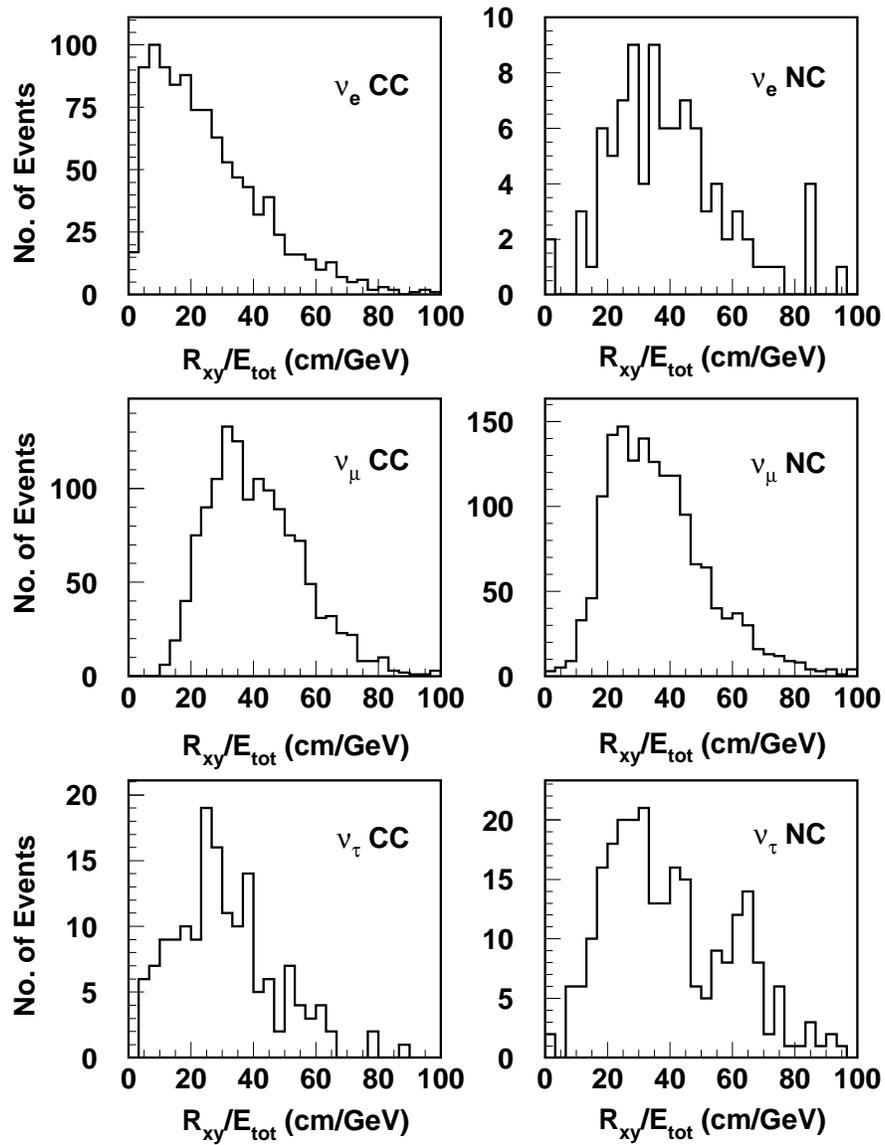,height=18.cm}}
\caption{The transverse event size 
         normalized to the total visible energy in the detector, 
         $\mathrm R_{xy}/\mathrm E_{tot}$,
         for various type of neutrino events. }
\label{fig:rxy}
\end{center}
\end{figure}

\begin{figure}[htbp!!!]
\begin{center}
\mbox{\epsfig{file=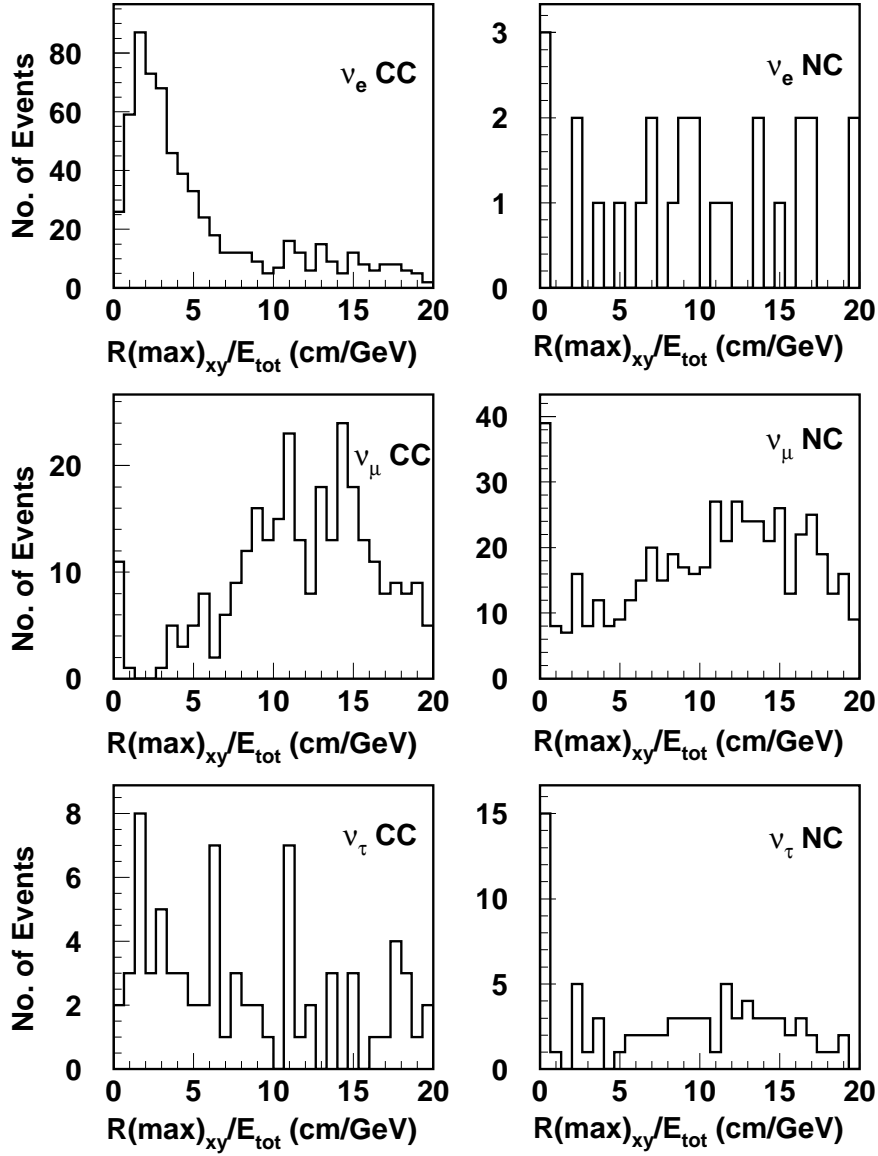,height=16.cm}}
\caption{The transverse event size at the shower maxima
normalized to the total visible energy in the detector, 
$\mathrm R^{max}_{xy}/\mathrm E_{tot}$
 for various type of 
neutrino events. The distribution of $\nu_e$ is different from all the others.}
\label{fig:rmaxxy}
\end{center}
\end{figure}

The neutrino CC events are identified by the following 5 variables:
$\mathrm E_{max}/\mathrm E_{jet}$, 
$\mathrm L_{shower}/\mathrm E_{jet}$, 
$\mathrm N_{tank}/\mathrm E_{jet}$,
$\mathrm R_{xy}/\mathrm E_{tot}$, 
and $\mathrm R^{max}_{xy}/\mathrm E_{tot}$, where 
$\mathrm E_{jet}$ is the jet energy, $\mathrm E_{tot}$ the total visible energy,
$\mathrm E_{max}$ the maximum energy in a cell,
$\mathrm L_{shower}$ the longitudinal length of the jet,
$\mathrm N_{tank}$ the number of cells with energy more than 10 MeV, 
$\mathrm R_{xy}$ the transverse event size and
$\mathrm R^{max}_{xy}$ the transverse event size at the shower maxima.
Fig.~\ref{fig:emax}-\ref{fig:rmaxxy} 
shows these variables for all
different neutrino flavors. 
It can be seen that $\nu_e$ CC events can be selected 
with reasonable efficiency and moderate backgrounds.

Table~\ref{tab:eff} shows the final results from this pilot Monte Carlo study.
For $\nu_e$ and $\nu_{\mu}$ 
events, $\nu_{\tau}$ CC events are dominant backgrounds, while for $\nu_{\tau}$,
the main background is $\nu_e$. It is interesting to see that this detector can 
identify $\nu_{\tau}$ in a statistical way. Similar results are obtained for a 
detector with 0.5m water tanks without RPCs. 
These results are similar to or better than those from water \C image 
detectors\cite{J2K}
and iron calorimeters\cite{wai2}. 
we would like to point out that using 
sophisticated jet reconstruction algorithm, shower shape analysis
and neural network,  
better results are expected.

\begin{table}[htb!]
\begin{center}
\begin{tabular}{|l|c|c|c|}
\hline
      & $\nu_e$ & $\nu_{\mu}$ & $\nu_{\tau}$  \\
\hline
CC Eff.             & 30\%  & 53\%      & 9.3\%  \\
\hline
$\nu_{e}$ CC        &  -    & $>$1300:1 & 3:1    \\ 
$\nu_{e}$ NC        & 166:1 &  665:1    & 60:1   \\
$\nu_{\mu}$ CC      & 700:1 &   -       & 270:1    \\
$\nu_{\mu}$ NC      & 92:1  & $>$6000:1 & 39:1   \\
$\nu_{\tau}$ CC     & 20:1  & 12:1      &  -   \\
$\nu_{\tau}$ NC     & 205:1 & 1100:1    & 61:1  \\
\hline
\end{tabular}
\caption{Results from Monte Carlo simulation: Efficiency vs background rejection 
power for different flavors.}
\label{tab:eff}
\end{center}
\end{table}

\begin{figure}[htbp!!!]
\begin{center}
\mbox{\epsfig{file=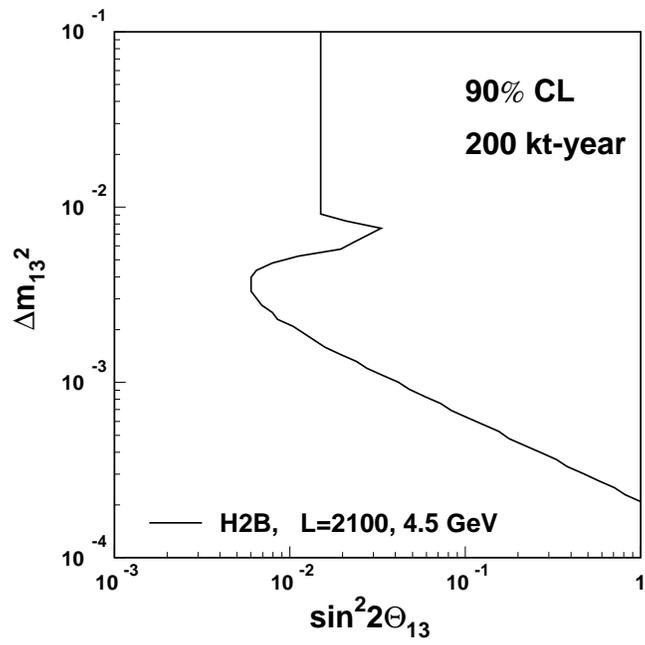,height=10.cm}}
\caption{Sensitivity to $\sin^22\theta_{13}$ at H2B. }
\label{fig:sens13}
\end{center}
\end{figure}

Using Table~\ref{tab:eff}, we can explore the sensitivity of our
detector to various quantities, such as  $\sin^22\theta_{13}$, CP phase term, 
etc. Fig.~\ref{fig:sens13} shows the sensitivity to $\sin^22\theta_{13}$ 
for 2 years running of a 100 kt detector.  If systematic errors can be 
controlled so that it does not dominate, we can reach 
$\sin^22\theta_{13} \sim 0.006$, corresponding to $\theta_{13} \sim 2.2^\circ$. 

\subsection{Other physics opportunities}

The detector we proposed above has an extreamly large mass, very good 
energy and angular resolution, hence a rich physics programs other
than long baseline neutrino oscillation experiments can be studied.
We discuss briefly other physics opportunities in the following, obviously
more detailed studies are needed. 
Depending on the over burden of the detector, some or all of the following 
incomplete list of physics topics
can be explored:

\begin{itemize}
{\setlength\itemsep{-0.6ex}
\item{\bf atmospheric neutrinos} A large data sample
of  atmospheric neutrinos with very good energy and angular resolution can be 
collected. We expect to improve the results obtained by Super-K.

\item {\bf Neutral current cross section} The charge current interaction
cross section is well-know, but there is very little data on neutral
current cross sections.  The BAND detector can contribute this area.  
\item{\bf supernova} If we can push the detection threshold down to about 10-15 MeV, 
this detector can observe supernovae at distances from us 
up to hundreds of kpc. 

\item{\bf primary cosmic-ray composition} By measuring multiple muons, the primary 
cosmic-ray composition, dominated either by heavy nuclei
or light nuclei, can be determined.

\item{\bf UHE cosmic rays} With an over burden of a few hundreds meters of water
equivalent, muons with energy above 50 GeV can be detected. This is an ideal
region for studying the composition of primary cosmic rays beyond
the "knee" and the details of hadronic interactions in the fragmentation region.

\item{\bf searches for dark matter} WIMP's as cold dark matter can annihilate in 
the core of the sun and the earth. By looking at excess of muons from the core of  
the sun and the earth, we can search for signs of
dark matter. For a few 100 kt detector, the allowed region from 
the DAMA experiment can be accessed.
 
\item{\bf searches for monopoles} Monopoles are predicted by many theories beyond 
the Standard Model. It can be
searched for by  looking at slow moving particles with high dE/dx in our detector. 
A few orders of magnitude improvement over results from MACRO are expected.

\item{\bf  searches for point sources} High energy neutrinos and muons from point 
sources  other than galaxies can provide important information about the evolution 
of the universe and the acceleration mechanism of cosmic-rays. Our detector can 
contribute significantly in this respect.

\item{\bf   searches for exotic particles} The history of physics is full of 
consummated surprises. ``Unexpected'' and ``exotic''  particles may be detected 
in our detector, such as fractionally charged particles.}
\end{itemize}

\newpage



\begin{thebibliography}{99}
\bibitem{okamura} A different approach of looking at the physics of the 
2100 km oscillation length in comparison with other
distances can be found in Ref.~\cite{kekkyoto}.
\end{thebibliography}
\end{document}